\documentclass[12pt]{iopart}

\usepackage{graphicx}
\usepackage{iopams}

\begin{document}
\title{Charmonium interaction in nuclear matter at FAIR}

\author{Partha Pratim Bhaduri}
\address{Variable Energy Cyclotron Centre, HBNI, 1/AF Bidhan Nagar, Kolkata 700064}
\ead{partha.bhaduri@vecc.gov.in}

\author{Michael Deveaux}
\address{IKF, Goethe University Frankfurt, 60438 Frankfurt/M, Germany}
\ead{deveaux@physik.uni-frankfurt.de}

\author{Alberica Toia}
\address{CBM Department, GSI, 64291 Darmstadt, Germany}
\address{IKF, Goethe University Frankfurt, 60438 Frankfurt/M, Germany}
\ead{a.toia@gsi.de}


\vspace{10pt}
\begin{indented}
\item[]December 2017
\end{indented}

\begin{abstract}
 We have studied the dissociation of $J/\psi$-mesons in low energy proton-nucleus ($p+A$) collisions in the energy range of the future SIS100 accelerator at Facility for Anti-proton and Ion Research (FAIR). According to the results of our calculations, various scenarios of $J/\psi$ absorption in nuclear matter show very distinct suppression patterns in the kinematic regime to be probed at FAIR. This suggests that the SIS100 energies are particularly suited to shed light on the issue of interaction of $J/\psi$ resonance in nuclear medium.

\end{abstract}

%% *******************************************************
\section{Introduction}
\label{intro}
%% *******************************************************
The observation of $J/\psi$ suppression in relativistic heavy ion collisions is considered as an evidence for the formation of quark-gluon plasma~\cite{MS,Vogt,Satz,Kluberg}. However, a considerable amount of $J/\psi$ suppression is also observed in proton-nucleus ($p+A$) collisions and commonly attributed to the dissociation of the meson in the cold nuclear matter (CNM) of the target nucleus~\cite{Satz,Kluberg}. A precise understanding of this so-called ``normal'' suppression is crucial to establish a robust baseline, with respect to which one can isolate the ``anomalous'' suppression pattern, specific to the dense QCD medium produced in heavy-ion collisions. Over a past few decades, $J/\psi$ production in proton-nucleus ($p+A$) have been studied extensively at several different fixed target experiments, in the beam energy range of $E_b = 158 - 920$ GeV~\cite{NA3,E772,NA38,NA50-200,NA50-400,NA50-450,E866,HERAB,NA60} and for a variety of nuclear targets. A common practice to estimate the cold nuclear matter effects is to fit the experimental results via the effective length traversed by the $J/\psi$ in the nuclear matter, calculated within Glauber model~\cite{Khar-Glauber}. The parameter $\sigma_{abs}^{J/\psi}$ extracted from the data quantifies the overall nuclear dissociation effects~\cite{Kluberg}. %If the initial state modification of the parton densities are explicitly taken into acount, $\sigma_{abs}^{J/\psi}$ siginifies the final state absorption cross section, exclusively due to the break up of the $c\bar{c}$ pairs in the pre-resonance or resonace stage. The magnitude of the cross section is in turn sensitive to the behavior of the opted parton distribution in the corresponding kinematic region~\cite{Lourenco}. 
The extraction of $\sigma_{abs}^{J/\psi}$ from $p+A$ data collected at SPS by the NA50 and the NA60 Collaborations revealed a significant beam energy dependence of the absorption cross section, with larger $\sigma_{abs}^{J/\psi}$ at lower beam energy~\cite{NA60}. This observation was in line with predictions discussed in~\cite{Lourenco}. %, prior to the $p+A$ runs performed by the NA60 Collaboration. 
 Within the Glauber model framework, the authors of Ref.~\cite{Lourenco}, analyzed the data on $J/\psi$ production cross sections measured in $p+A$ collisions in fixed target experients, with proton beam energies from 200 to 920 GeV and in $d+Au$ collisions at RHIC, at $\sqrt{s_{NN}}= 200$ GeV. Several sets of parton distributions with and without nuclear modifications were explicitly employed to account for the  initial state effects. The magnitude of  the final state absorption cross section, $\sigma_{abs}^{J/\psi}$, is found to be sensitive to the behavior of the opted parton distribution in the corresponding kinematic region. Moreover the results revealed a significant dependence of $\sigma_{abs}^{J/\psi}$, on the kinematics of the $J/\psi$ and on the beam enrgy of collision, which were extraploated to estimate the expected level of absorption in $p+A$ collisions at 158 GeV.

The production of charmonium in nuclear collisions, requires a certain formation time. The related formation length in the target nucleus rest frame depends on the relative velocity of the $c\bar{c}$ pair and may exceed the diameter of the nucleus. Slow $c\bar{c}$ pairs would form physical resonances inside the target nucleus while fast pairs form this resonance only in vacuum. It is likely that fully formed resonances show different interactions with the nuclear medium than the evoling $c\bar{c}$ pairs. The velocity of the produced $c\bar{c}$ pairs depends on the velocity of the beam proton w.r.t the target nucleus and the kineamtic domain explored by the charmonium production. In~\cite{SatzPaper1}, the authors analysed the kinematic regimes attainable for $J/\psi$ and $\psi'$ production in 160 GeV $p+A$ collisions. 
Selecting fully formed resonances is a prerequisite for observing differences in the interactions of different charmonium states with the nuclear medium.
Their results indicated that it is required to probe the phase space region $x_F \le -0.45$, where $x_F$ denotes the Feynman scaling variable. In this region, the produced $c\bar{c}$ pairs at 160 GeV, are slow enough to be produced inside the target nucleus. This eventually led to distinguishably different suppresion patterns by $J/\psi$ and $\psi'$ resonances due to their different binding energies. Even though NA60 Collaboration took data in 158 GeV $p+A$ collisions, their measurements~\cite{NA60} were confined in the positive hemisphere over a rapdity range $0.28 < y_{cms} <0.78$, which corresponds to the $x_F$ domain $0.1 < x_F < 0.3$. Also no $\psi'$ data are available from NA60 Collaboration due to limitation in statistics.

Up till now there is no significant measurement of the cold nuclear matter effects on charmonium production in $p+A$ collisions below 158 GeV. The situation changes with the appearance of the Compressed Baryonic Matter (CBM) experiment at FAIR~\cite{cbm-paper}. The CBM detector set up at SIS100 is suited to measure charmonia and open charm hadrons in $p+A$ collisions, at proton beam energies from $15-30~\rm GeV$, thanks to its unprecedented rate capability. These measurements will be highly interesting to investigate the potential issue of $J/\psi$ interaction in nuclear medium. In the present article, we discuss the kinematics of $J/\psi$ production in 15 and 30 GeV $p+Au$ collision systems, as available at the SIS 100 accelerator at FAIR. We rely on the formulation of production kinematics developed in~\cite{SatzPaper1}. The interesting question on charm propagation in nuclear matter that can be addressed via charmonium measurements at FAIR was first qualitatively triggered in~\cite{cbm-satz}. This was certainly a call for a more detailed study. We calculate in quantitaive details the differential distribution of $J/\psi$ production cross sections in $p+Au$ collisions.
Moreover, we discuss if the $J/\psi$ production cross-section observed in $p+A$ collisions at SIS100 is a suitable probe to distinguish different models for charmonium dissociation. To do so, we estimated this cross-section for different models for nuclear dissociation of $J/\psi$. The resultant suppression patterns are found to be distinguishably different, giving us the opportunity to probe the mechanism of charmonium dissociation in nuclear medium.  

\section{Theoretical formulation}
\label{sec:1}
\subsection{Kinematics}
\label{sec:2}

\begin{figure*}[t]
\center
\resizebox{0.95\textwidth}{!}{%
  \includegraphics{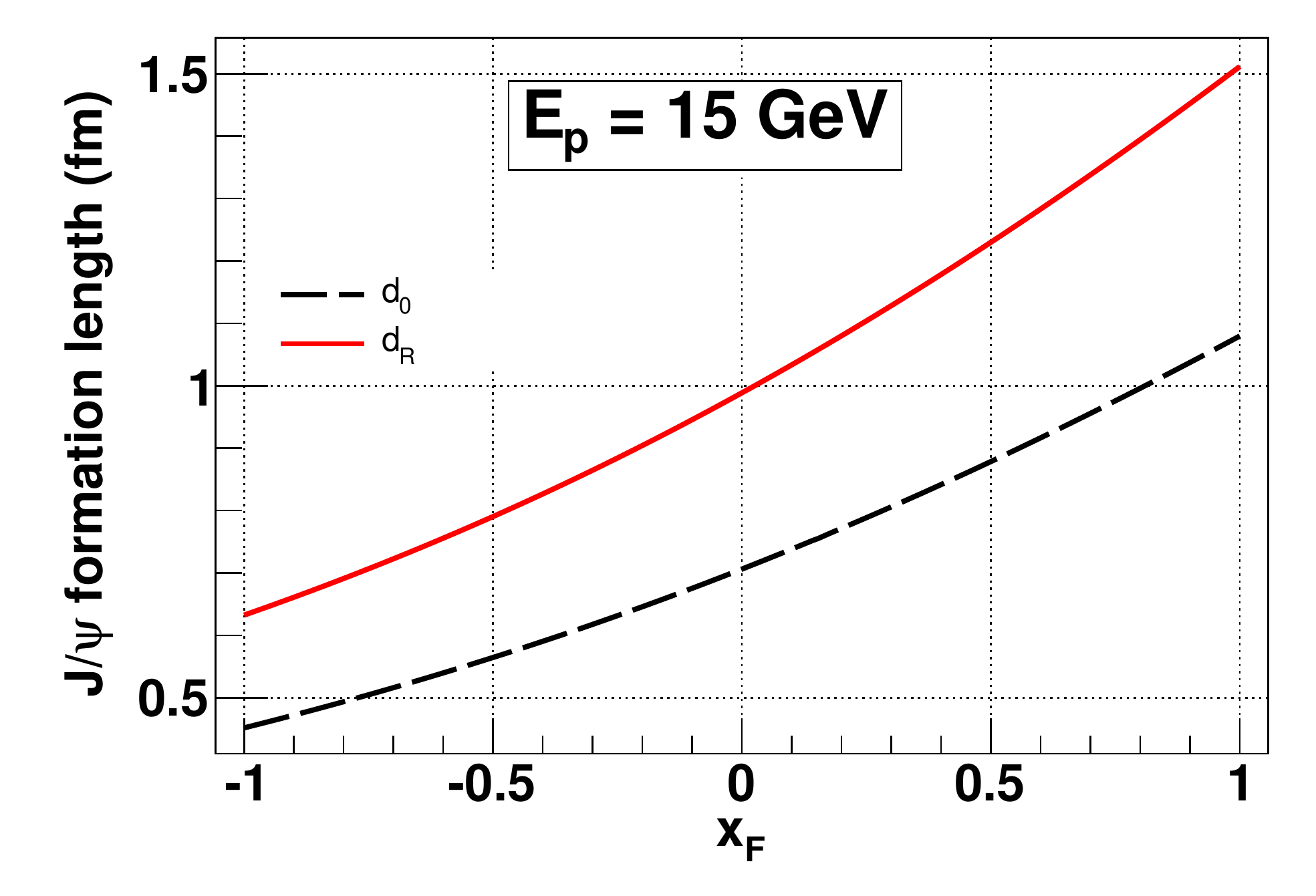}
   \includegraphics{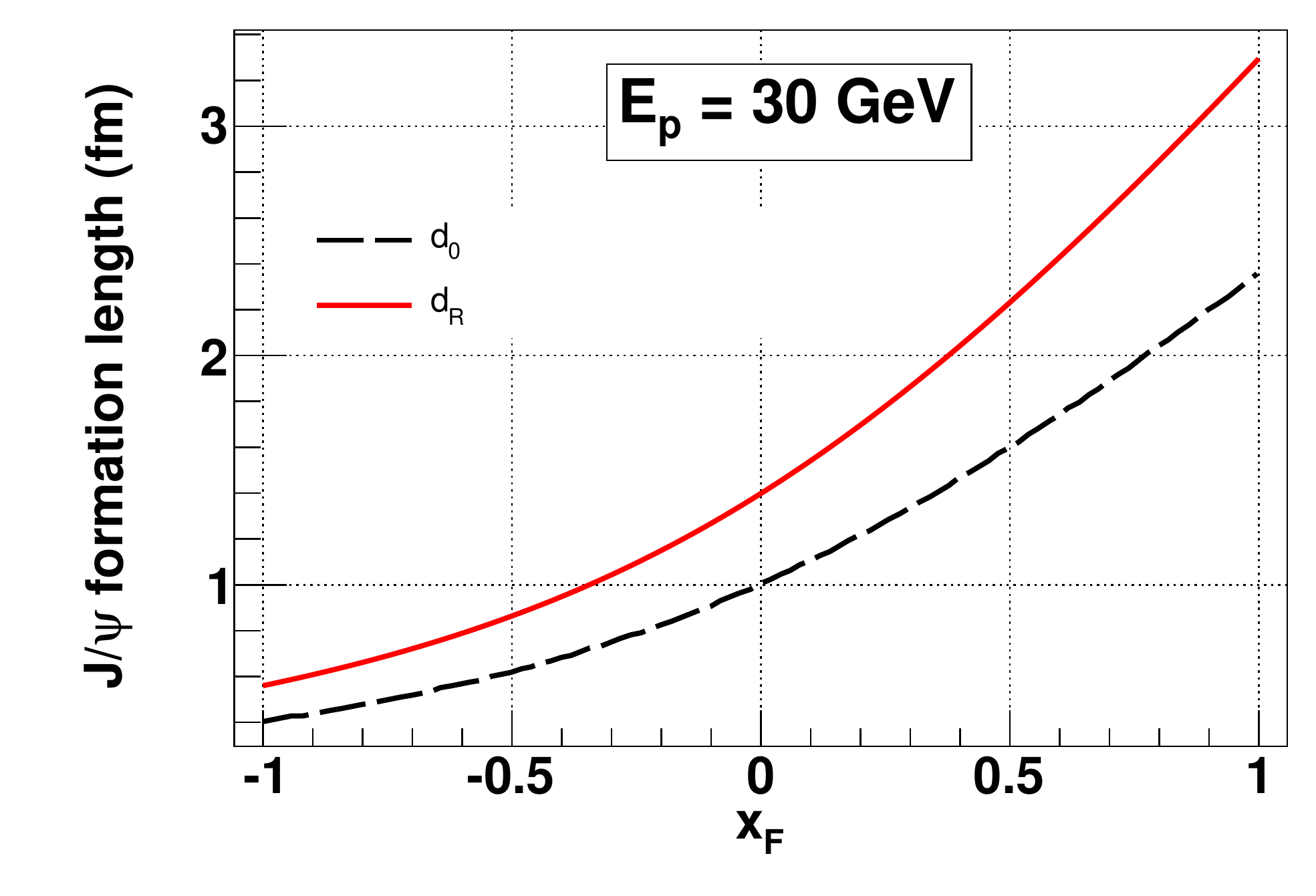}
}
\caption{$x_F$ dependence of the formation length of $J/\psi$ mesons, in the laboratory frame, in 15 GeV (left panel) and 30 GeV (right panel) $p+A$ collisions. At 15 GeV beam energy, the resonance formation lengths remain less then 2 fm, even for the fastest mesons. At 30 GeV, formation length is longer than 2 fm, implying the propagation of expanding colorless $c\bar{c}$ states beyond $x_{F}  \simeq 0.4$.}
\label{Fig1}
\end{figure*}
In the literature, $J/\psi$ production in hadronic collisions is usually considered as a factorizable two step process. The first step is the production of a color octet $c\bar{c}$ pair that can be described by perturbative QCD (pQCD). This is followed by the non-perturbative formation of the color singlet resonance, which requires a finite time (see for example Ref.~\cite{Anton} for an up-to-date review of the quarkonium production up to the LHC energies). In the $c\bar{c}$ rest frame, color neutralization occurs at a time scale of $\tau_{0} \simeq 0.25$ fm~\cite{SatzPaper1}.
%this time is quoted as $\tau_{0} \simeq 0.25$ fm in the $c\bar{c}$ rest frame, as calculated from the energy of the evaporated gluons in confined matter. 
Physical resonances with appropriate size and quantum numbers are believed to take even longer time to form. In our following calculations, we would adopt a value $\tau_{R} \simeq 0.35$~\cite{Karsch} fm for intrinsic formation time. However the choice of the resonance formation time is to some extent arbitrary and alternative estimates are available in literature~\cite{formation}. 

In $p+A$ collisions, to undergo the nuclear medium effects by the resonance itself, the charmonium states need to be formed inside the nuclear medium or even better before hitting a nucleon, apart from the one on which it is produced. The second condition is met if the resonance formation length in the laboratory frame, remains below the average distance of two nucleons in the core,  which is assumed to amount $\sim 2~\rm fm$ in our calculations. Formation length scales in the laboratory frame can be estimated as:
\begin{equation}
d_{0(R)}= \beta \gamma c \tau_{0(R)}=\frac{P_{L}}{M} \tau_{0(R)}
\label{Eq1}
\end{equation}
where $d_{0}$ and $d_{R}$ stand for the formation lengths of color singlet $c\bar{c}$  pair and of fully developed  resonance state respectively. The mass and momentum of the $c\bar{c}$ resonance state are denoted by $M$ and $P_{L}$, respectively. $\beta$ is the velocity of the state in the laboratory frame. The center-of-mass (CMS) momentum ($P_{CMS}$) of the resonance state amounts: 
\begin{equation}
P_{CMS} = \gamma_{CMS}P_{L} - \gamma_{CMS}\beta_{CMS}\sqrt{P_{L}^{2}+M^{2}}
\label{Eq2}
\end{equation}
where $\beta_{CMS}$ is the velocity of the CMS in the laboratory. %In this frame, the beam proton and the nucleus (including the N, which is to collide) impinge from opposite side and their velocity is identical. 
The maximum momentum of a $J/\psi$ meson in the CMS frame in an elementary reaction, like $p+N \rightarrow J/\psi+p+N$, occurs if the two nucleons travel both in the one direction opposite to that of $J/\psi$. If the two nucleons travel in forward direction and $J/\psi$ in backward direction, then in the laboratory frame $J/\psi$ would be emitted with the minimum possible momentum (slowest $J/\psi$). The maximum CMS momentum ($P_{max}$) can be obtained from the relation:
\begin{equation}
s= {\left(\sqrt{P_{max}^2+M^2} + \sqrt{P_{max}^2+4m^2}\right)}^2
\label{Eq3}
\end{equation}   
where $s$ and $m$ denote the square of CMS energy and the nucleon mass respectively. Solving Eq.\ref{Eq3} one obtains
\begin{equation}
P_{max} = \sqrt{\left(\frac{s+4m^2-M^2}{2\sqrt{s}} \right)^2 - 4m^2}
\label{Eq4}
\end{equation}

Concerning the interaction between the charmonium and the nuclear medium, three different kinematical regimes can be distinguished. In a first case, called ``color octet'' region, the $c\bar{c}$ pair penetrates the nuclear core before forming a resonance and the resonance is formed in vacuum. In the so-called ``resonance region'' the resonance is fully developed before the $c\bar{c}$ pair hits a nucleon. In the ``transition region'', the formation occurs while the $c\bar{c}$ penetrates the core.

\begin{figure*}[t]
\center
\resizebox{0.95\textwidth}{!}{%
\includegraphics{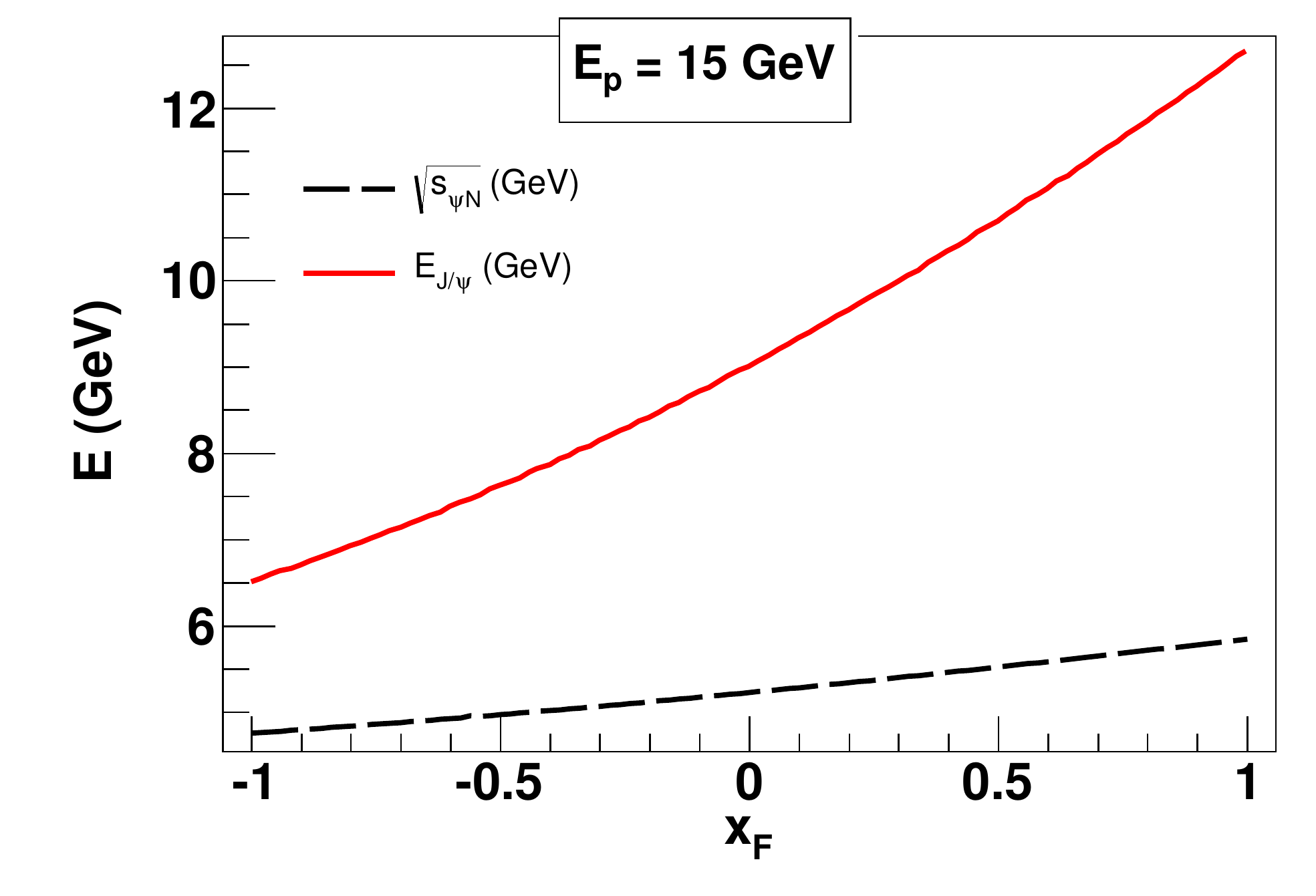}
\includegraphics{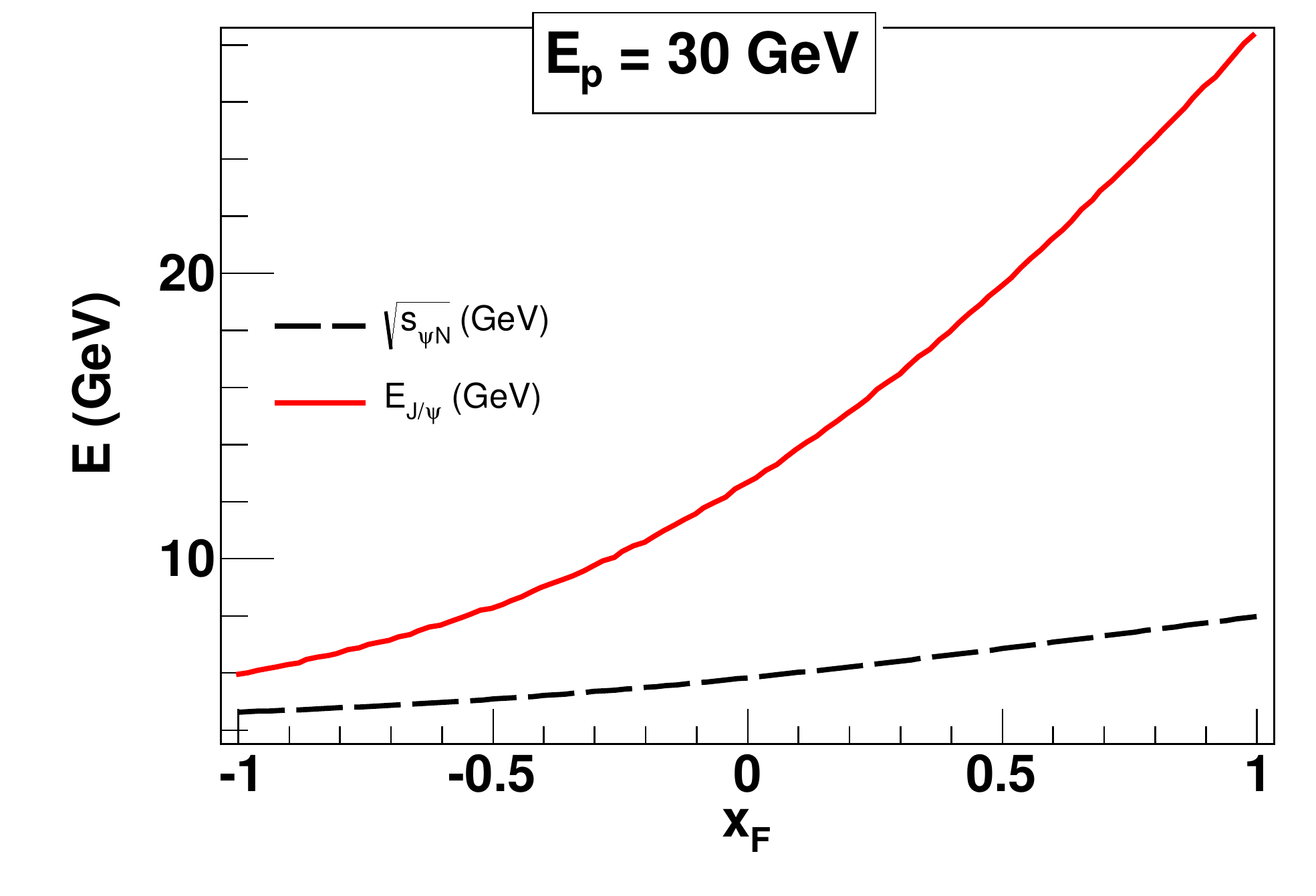}
}
\caption{$x_F$ dependence of the energy of the $J/\psi$ mesons in the laboratory frame, and the corresponding CMS energy of $J/\psi + N$ interaction, in 15 GeV (left panel) and 30 GeV (right panel) $p+A$ collisions.}
\label{Fig2}
\end{figure*}

The inelastic $J/\psi-N$ dissociation cross section ($\sigma_{J/\psi}$) can be measured by comparing the yield in $p+A$ with the one observed in $p+p$ collisions. The survival probability $S_{J/\psi}$ for fully formed $J/\psi$ mesons depend on the average path length inside the target nucleus ($L_A$) as follows:
%, preferably in the ``resonance region'', as a function of the average path length ($L_{A}$) inside the target. The corresponding survival probability ($S_{J/\psi}$)  is given by:
\begin{equation}
S_{J/\psi}= \exp(- n_0 \sigma_{J/\psi} L_{A})
\label{Eq5}
\end{equation}
where $n_0=0.15~\rm fm^ {-3}$ is the saturation nuclear density. Following prescription given in~\cite{Wongbook}, we have used $L_A \simeq {3 \over 4} R_A$ ($R_A$ being nuclear radius) for heavy nuclei. In the ``transition region'', the dissociation cross section changes while the pair is being formed. Based on a classical color dipolar approximation~\cite{SatzPaper1}, the dissociation cross section of an evolving color neutral $c\bar{c}$ pair can be parametrized as:

\begin{equation}
\sigma (d) = \sigma_{0} \left(\frac{d}{\bar L} \right)^{2} 
\label{Eq6}
\end{equation}
Here, $d$ denotes the instantaneous size of an expanding $c\bar{c}$ pair and $\sigma_0$ is the dissociation cross section for fully developed resonances by nucleons inside the target. $\bar{L}$ stands for the effective distance travelled until full resonance formation.  %A more rigorous treatment of the evolving color transparent resonances, including quantum effects can be found in~\cite{Gerland}. 
Using the parameterization would modify the survival probability given in Eq.~\ref{Eq5} as 
\begin{equation}
S_{J/\psi} = \exp\left(-n_{0}\sigma_{0}\left[L_A - {2 \over 3}\bar{L}\right]\right)
\label{Eq7}
\end{equation}
which remains valid until the resonance is fully formed. Note that there are other parametrizations available for modelling the dissociation of an expanding $c\bar{c}$ pair~\cite{Arleo}. As the extent of the transition region (where this scenario would be operative) is not large at CBM energies, we refrain from using them.

\subsection{Scenario at FAIR}
\label{sec:3}
\subsubsection{Determination of the kinematic region}
\label{sec:4}
 The kinematic threshold for charmonium production in $p+p$ collisions amounts \mbox{$E_{th}^{J/\psi} \simeq 12.2$ GeV} for $J/\psi$, $E_{th}^{\psi'} \simeq 15.6$ GeV for $\psi'$, and $E_{th}^{\chi_{c}} \simeq 14.6$ GeV for $\chi_{c}$. The SIS100 synchrotron at FAIR will provide proton beams up to 30 GeV. This allows to produce all three particles near but above threshold. However, experiments will be presumably restricted to $J/\psi$ due to the low yield of the other mesons. 

At a beam energy of 15 GeV, the maximum momentum of a $J/\psi$ in the CMS frame of the initial collision amounts $|P_{max}| = 1.3$ GeV. The momenta in the laboratory frame range from 
$P_{x_F=-1} = 5.58$ GeV to $P_{x_F=1} = 13.34$ GeV. This corresponds to a rapidity coverage of $1.35 < y_{J\psi}^{Lab} < 2.17$. A $J/\psi$ produced at rest in the CMS frame would fly with a monemtum of $8.73$ GeV in the laboratory frame. Singlet as well as resonance formation lengths in the laboratory frame are displayed in Fig.~\ref{Fig1} and remains below 2 fm over the full range of $x_{F}$. Therefore, all the $J/\psi$ are formed in the ``resonance region''. At a beam energy of 30 GeV, we find $|P_{max}| = 2.94$ GeV.  In the laboratoy frame, the momentum range spans from  
$P_{x_F=-1} = 4.94$ GeV to $P_{x_F=1} = 29.1$ GeV, which covers a rapidity range $1.25 < y_{J\psi}^{Lab} < 2.94$. At this energy, the momentum of a $J/\psi$ produced at rest in the CMS frame would be $12.3$ GeV. The corresponding formation lengths shown in Fig.~\ref{Fig1} suggest that the ``resonance region'' ranges up to $x_{F}  \simeq 0.4$ and that a ``transition region'' is found at higher $x_{F}$. Once formed, the $J/\psi$ mesons may be absorbed (dissociated) by processes like $J/\psi + N \rightarrow \Lambda_{C} + \bar{D}$ or like $J/\psi + N \rightarrow D + \bar{D} + N$, while travelling inside the nuclear medium. As this dissociation is endothermic, its cross section depends most likely on the CMS energy ($\sqrt{s_{{\psi}N}}$) of the $J/\psi + N$ collision. This energy is plotted in Fig.~\ref{Fig2} along with the energy of the $J/\psi$ mesons ($E_{J/\psi}$) in the laboratory frame.$\sqrt{s_{{\psi}N}}$ spans from 4.76 to 5.9 GeV for 15 GeV beam energy, whereas at 30 GeV, it covers a range from 4.6 to 8 GeV. The threshold CMS energy required for the $\Lambda_{c}\bar{D}$ production is $4.1$ GeV, the one of the  $D\bar{D}N$ channel is $4.7$ GeV. Therefore, experiments carried out at FAIR would conceptually be able to map out the cross sections close to the related kinematical thresholds. 

\subsubsection{Estimation of dissociation cross sections}
\label{sec:5}
\begin{figure}[!h]
\center
%\caption{Variation of $J/\psi$ dissociation cross section by nucleons as obtained from perturbative QCD in 15 GeV $p+A$ collisions.}
\resizebox{0.5\textwidth}{!}{%
  \includegraphics{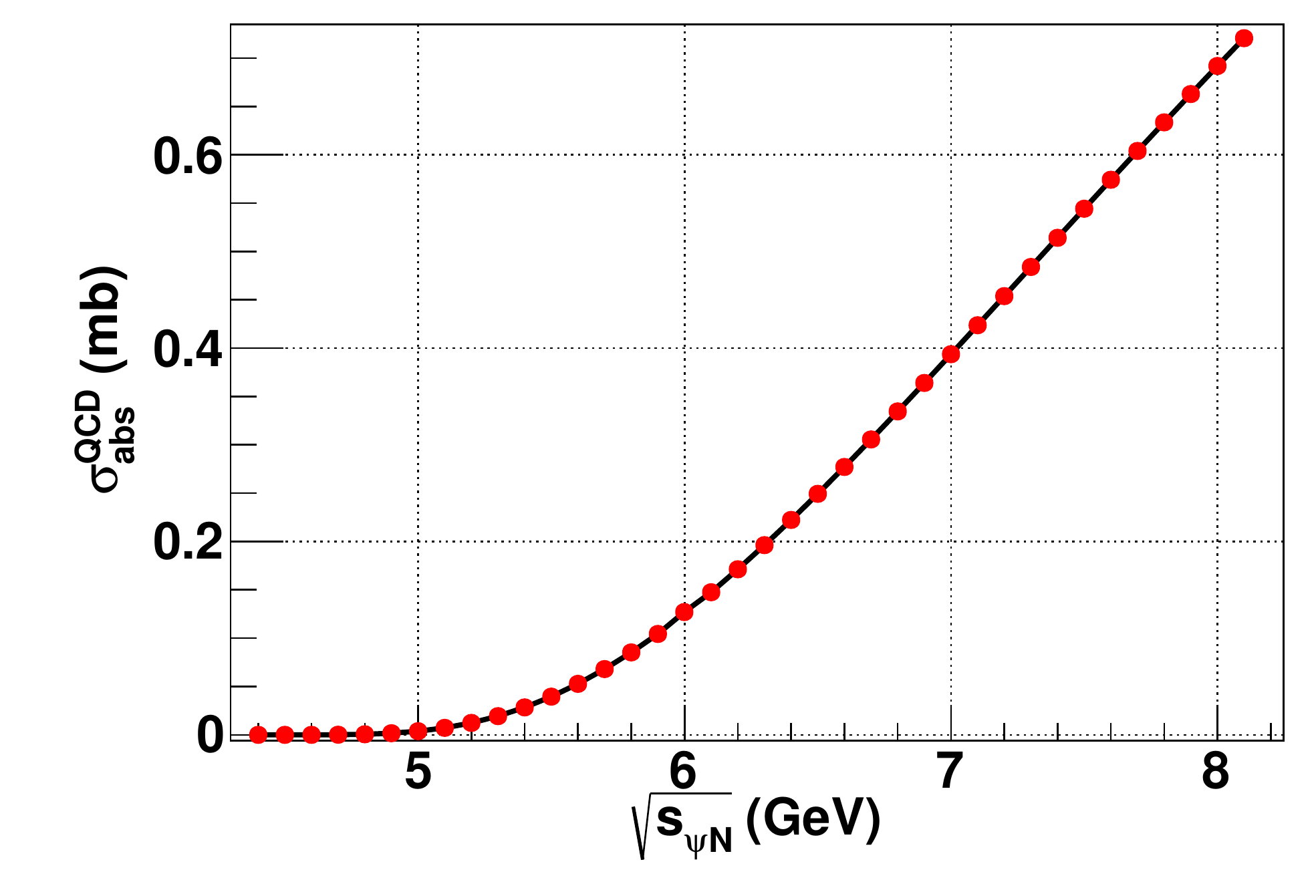}
}
\caption{Variation of $J/\psi$ dissociation cross section by nucleons as obtained from pQCD in $p+A$ collisions in the FAIR energy domain.}
\label{Fig3}
\end{figure}
Measuring the absorption cross section $\sigma_{abs}$ with the above mentioned approach requires that the absorption is sufficiently strong to modify the survival probability of $J/\psi$ significantly. As there is no direct experimental measurements on $\sigma_{abs}$, we study this question based on different theoretical estimates available in literature. 

Within geometric approach, $\sigma_{abs}$ is proportional to the square of the radius of the particular charmonium state.  
%Spatial sizes are generally calculated from the non-relativistic quarkonium models and thus depend on the shape of the input $c\bar{c}$ potential~\cite{Gerland1,Gerland2}. %Calculations based on the model of stochastic vacuum (MSV)~\cite{Dosch} result in non-perturbative $J/\psi + N$ cross sections obtained from the weighted average of the longitudinally and transversely polarized $J/\psi$ wave functions, which amount $\sigma_{abs} \simeq 4.4$ mb. 
Even though validity of asymptotic cross sections near threshold is not free from doubt, in our calculations we use $\sigma_{abs}^{Geo} \simeq 2.5$ mb for $J/\psi$ following~\cite{SatzPaper1}.
%It was adopted from~\cite{Hufner1987} and corresponds to a spatial size of $r_{J/\psi} \simeq 0.28$ fm. As argued in~\cite{SatzPaper1}, application of asymptotic cross sections to estimate the charmonium absorption in nuclear matter in low energy $p+A$ collisions is not free from doubt. The underlying strong assumption is that the interaction cross sections attain their full asymptotic values near threshold. 
\begin{figure} [!h]
\center
\resizebox{0.5\textwidth}{!}{%
\includegraphics{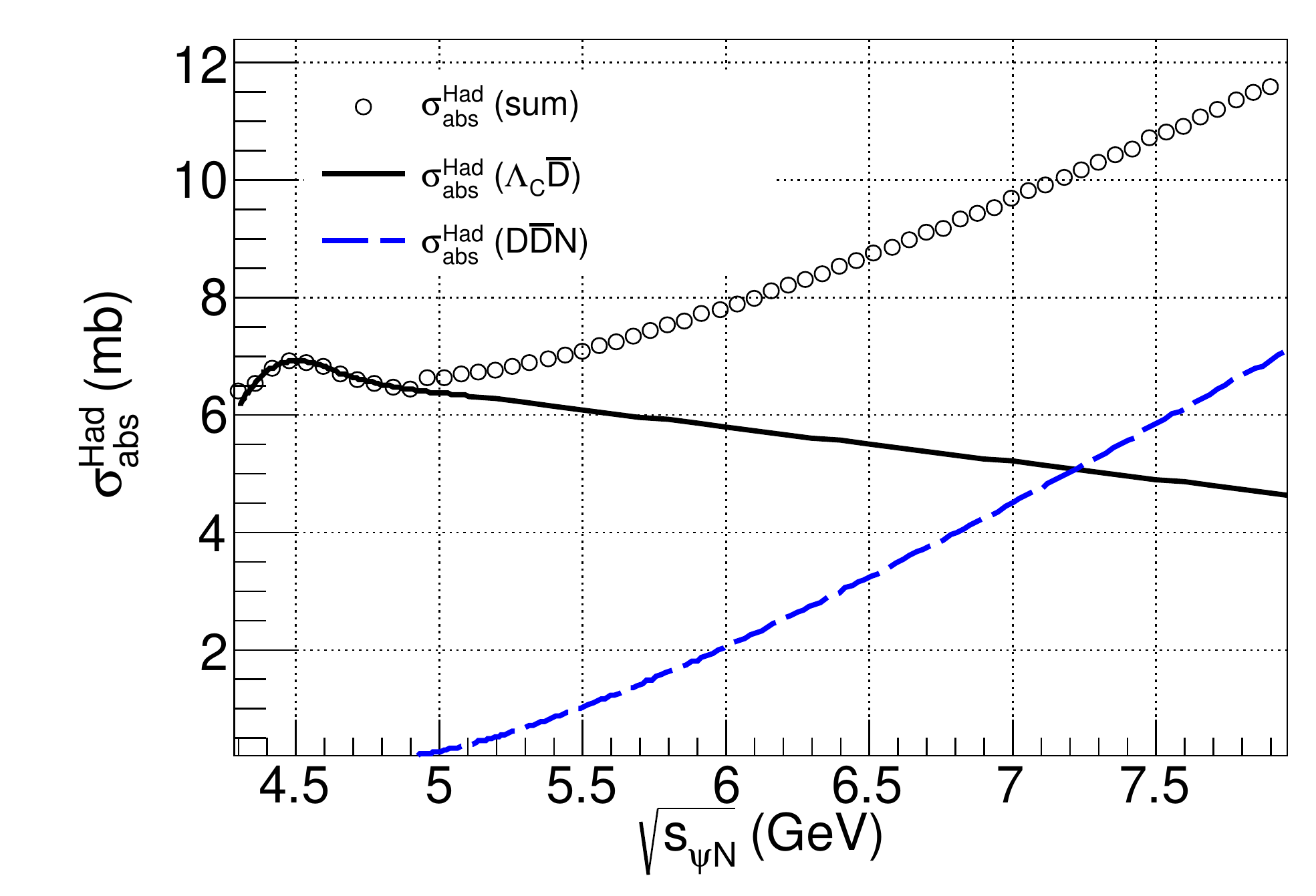}}
%\label{Fig4}
\caption{Variation of $J/\psi$ dissociation cross section by nucleons, as obtained from hadronic models in 15 GeV $p+A$ collisions. The effect of Fermi motion of the nucleons inside the nucleus is taken into account.}
\label{Fig4}
\end{figure}

The theoretical estimates of the dynamical dissociation cross sections can be broadly divided into two categories. The first approach is based on pQCD. 
%Due to small spatial size and large binding energy of $J/\psi$ mesons, only hard gluonic content of the colliding nucleons can cause the dissociation. The break up cross section is calculated within the ambit of short distance QCD using operator product expansion. 
The nuclear dissociation cross section within this framework using QCD sum rules was found to be approximately parametrized as~\cite{KharzeevPLB94}
\begin{equation}
\sigma_{J/\psi} \approx 2.5 {\rm mb} \times {\left[ 1- \left(\frac{2 M_{J/\psi} (m_N + \epsilon_{J/\psi} )}{({s_{{\psi}N}} - M^2_{J/\psi})} \right) \right]}^{6.5}
\label{Eq8}
\end{equation}
where, $m_N=0.94~\rm GeV$ is the mass of the target nucleon, $\epsilon_{J/\psi} = 0.64~{\rm GeV}$ the binding energy of the $J/\psi$. Since the bound nucleons do not contain a sufficient number of hard gluons, this cross section shows a large threshold damping. Over the energy range to be probed at FAIR, this cross section is around an order of magnitude smaller than the corresponding asymptotic values as evident from the Fig.~\ref{Fig3}. A more rigorous discussion of the pQCD inspired dissociation cross sections can be found in~\cite{KharzeevPLB99,SHLee}. The main caveat in this theory is that the $J/\psi$ is considered as a Coulombic bound state, which could only be applicable at very large charm quark mass limit of $m_{c} > 25$ GeV. However results from vector meson dominance (VDM) model were found to be in agreement with the short distance QCD calculations~\cite{Redlich}. 

\begin{figure*}[t]
\center
\resizebox{0.95\textwidth}{!}{%
\includegraphics{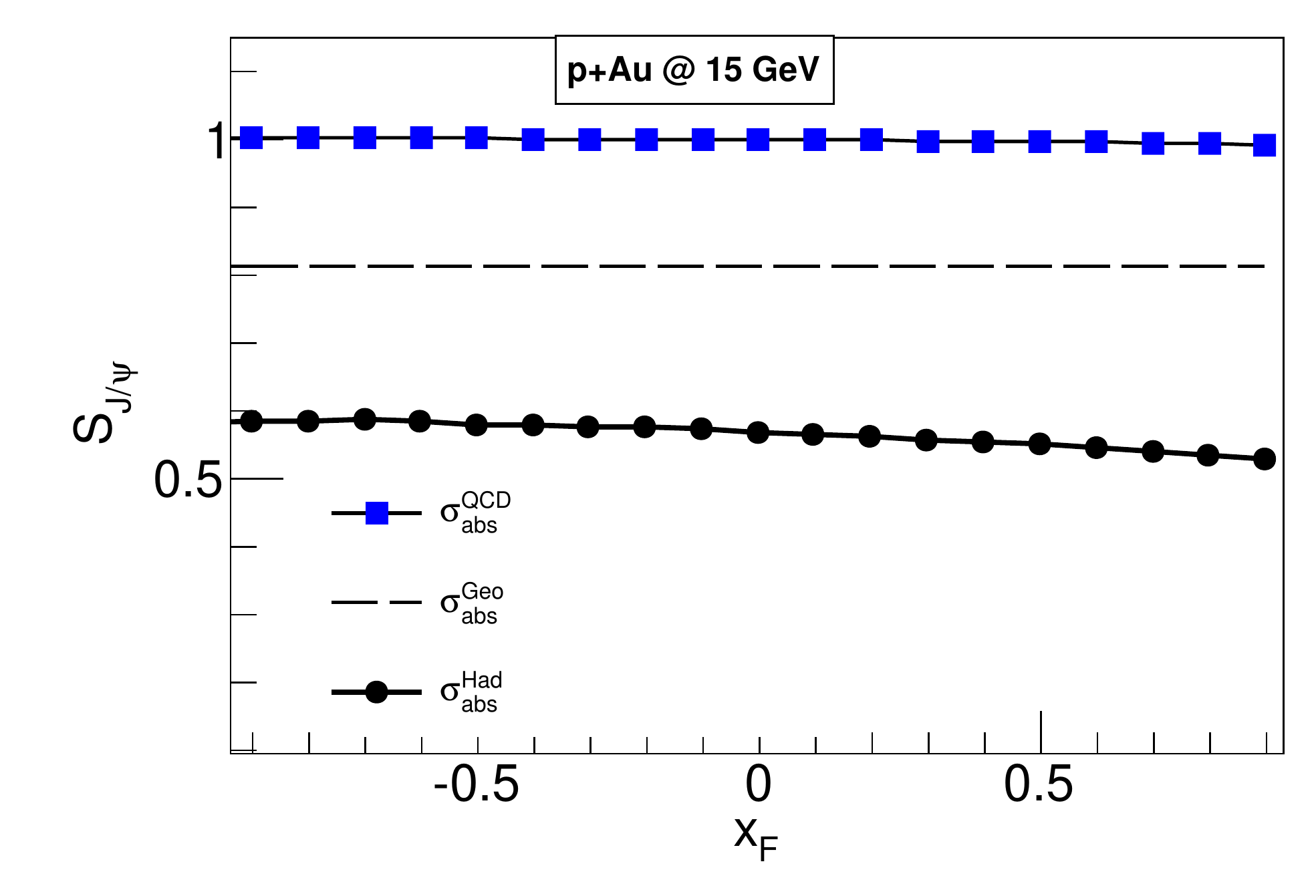}
\includegraphics{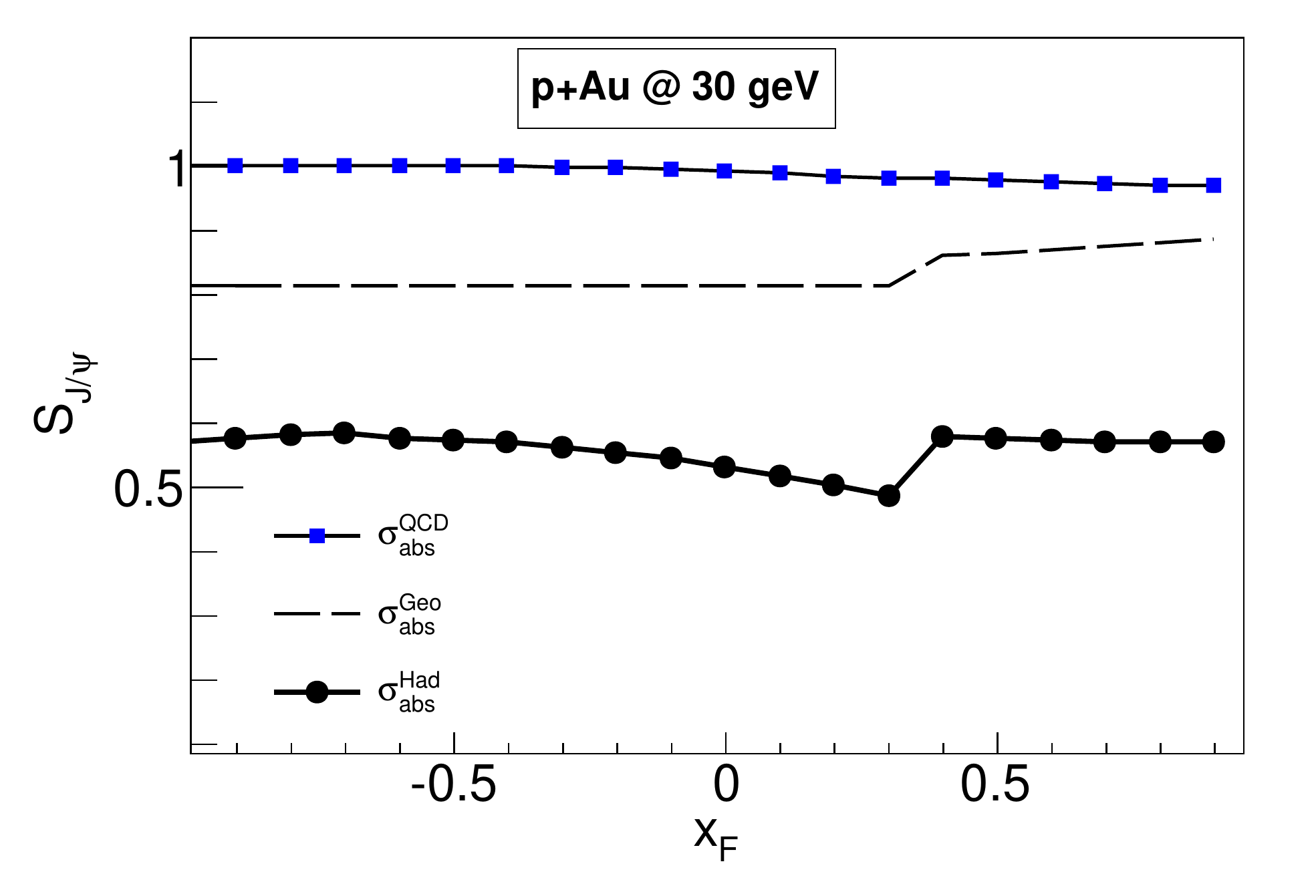}
}
\caption{$x_F$ dependence of the $J/\psi$ survival probability in $p+Au$ collisions at FAIR. The effect of feed down from excited states is not taken into account.}\label{Fig5}
\end{figure*}
The other approach is non perturbative and uses hadronic models based on quark exchange~\cite{Martins,Brane} or meson exchange~\cite{Muller,Haglin,Oset}. Due to a lack to experimental data, the total inelastic $J/\psi + N$ cross sections as predicted by these models spreads over a sizeable range. But a common feature of all these model is that the dissociation cross section peaks close to threshold, which stands in contrast to the pQCD predictions. To the best of our knowledge, the most recent effective theory calculations on $J/\psi$ interaction with nuclei appear in~\cite{Oset}. The dominant contribution to $J/\psi$ dissociation comes from $\Lambda_{C}\bar{D}$ channel, due to the lowest threshold. The total inelastic cross section peaks around $\sqrt{s_{{\psi}N}} = 4.46$ GeV which gets diluted once the cross section is averaged over the momenta distributions of the nucleons inside the nucleus. However the cross sections available in these calculations are restricted to a ${J/\psi}-N$ cms energy of up to 5 GeV. To extend this data to the full FAIR energy range of $\sqrt{s_{{\psi}N}} \simeq 4.5 - 8$ GeV, one has to complement the $J/\psi$ dissociation via $D\bar{D}N$ channel as well, which is not included in~\cite{Oset}. An estimate of the dissociation cross section via both, the $\Lambda_{C}\bar{D}$ and the $D\bar{D}N$ channel, are available in~\cite{Sibirstev}. %The cross section are calculated for both inelastic channels considering interactions via the $D$ and $D^*$ meson exchange.
Including the $D\bar{D}N$ channel leads to a monotonic rise of the total inelastic cross section with energy, which would otherwise descend beyond the $\Lambda_{C}\bar{D}$ threshold. However the total cross section computed for the $\Lambda_{C}\bar{D}$ channel amounts more than a factor of two less than the ones calculated by other authors~\cite{Haglin,Oset}. This is plausibly because the contact interaction is ignored, which seems to provide the dominating contribution at low energies. 
\begin{figure}%[!h]
\center
\resizebox{0.5\textwidth}{!}{%
  \includegraphics{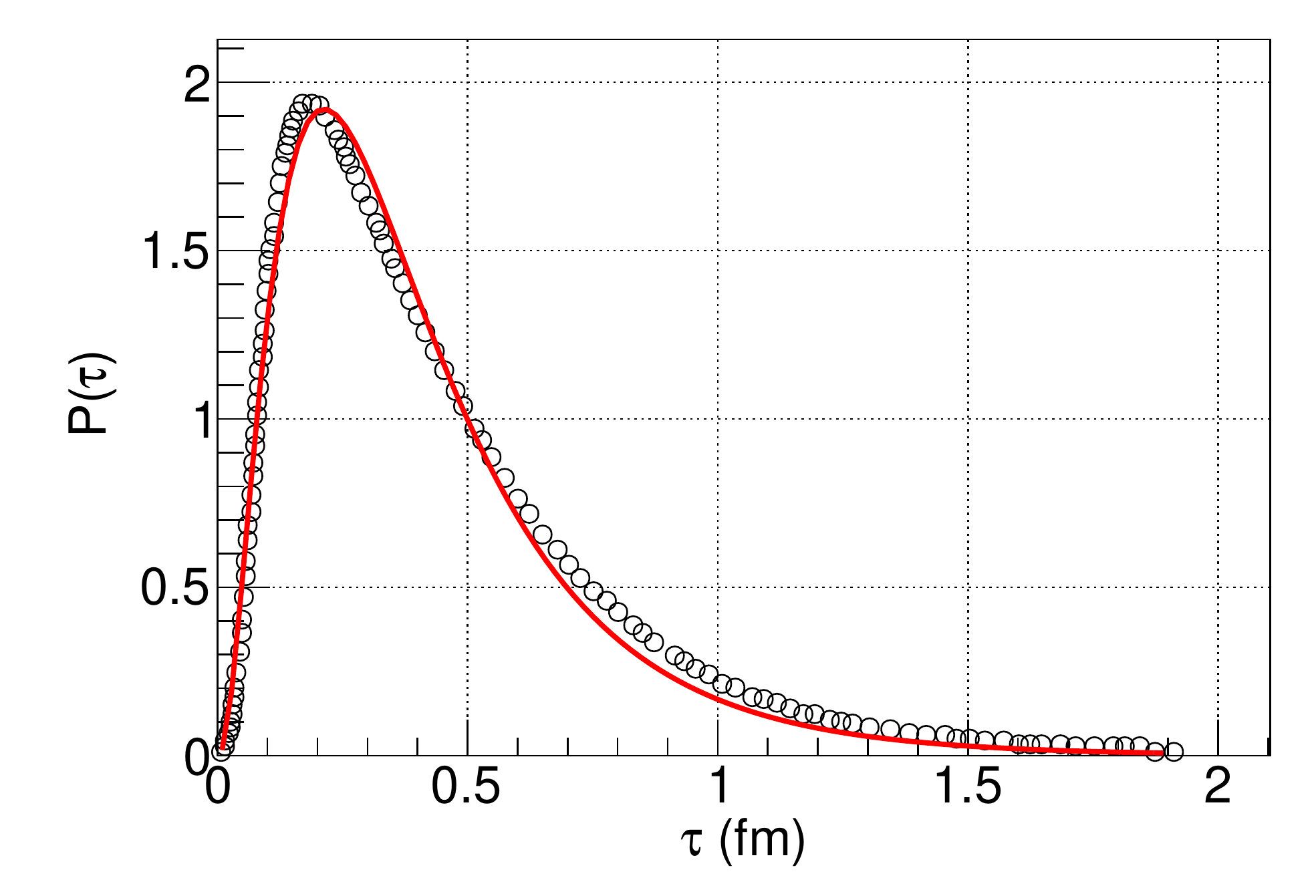}
}
\caption{Distribution of $J/\psi$ formation time. Black circles denote the data points from the original calculations and the red line corresponds to the fitted curve required for our modelling.}
\label{Fig7}
\end{figure}
In the absence of suitable calculations that include all relevant processes with appropriate interactions at FAIR, we estimate the energy dependence of the total inelastic cross section as follows: Up to $D\bar{D}N$ threshold, we take the energy dependence from~\cite{Oset}, where as beyond this the total cross section is the sum of the extrapolated cross section from~\cite{Oset} and the inelastic cross section from $D\bar{D}N$ channel as reported in~\cite{Sibirstev}. Since the choice of form factors in such calculations is no way unique, we took the cross section without form factor corrections. The energy dependence of the cross sections for the two processes and their sum as included in our calculations is shown in Fig.~\ref{Fig4}.  For consistency, we used the raw inelastic cross sections and made suitable extrapolations with linear functions where necessary. Hereafter, we calculated the average cross section accounting for the Fermi motion of the nucleons. %The Fermi momentum is evaluated with nuclear density $\rho = \rho_{0} = 0.17$ fm$^{-3}$ and comes out to be $p_{F} = 267$ MeV. 
The total average cross section shows a small bump around $\Lambda_{C}\bar{D}$ and then increases monotonically with energy due to growing contribution from the $D\bar{D}N$ channel. In the kinematic domain probed by FAIR, the dissociation cross sections from hadronic models is orders of magnitude larger than the pQCD inspired values. With increase in energy of the $J/\psi+N$ collisions, the difference in the cross sections from the two approaches gradually decreases~\cite{Sibirstev}. 

\begin{figure*}[t]
\center
\resizebox{0.95\textwidth}{!}{%
\includegraphics{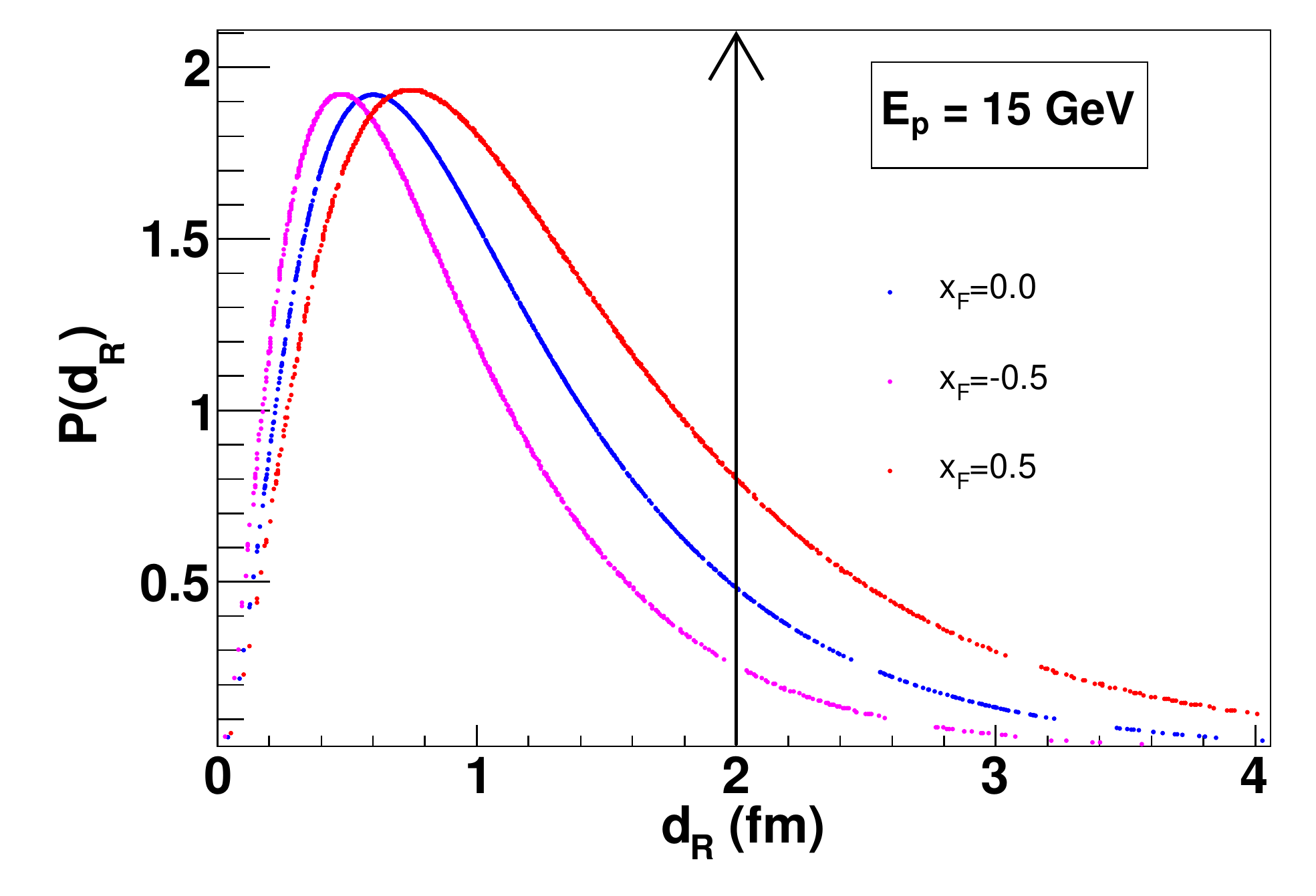}
\includegraphics{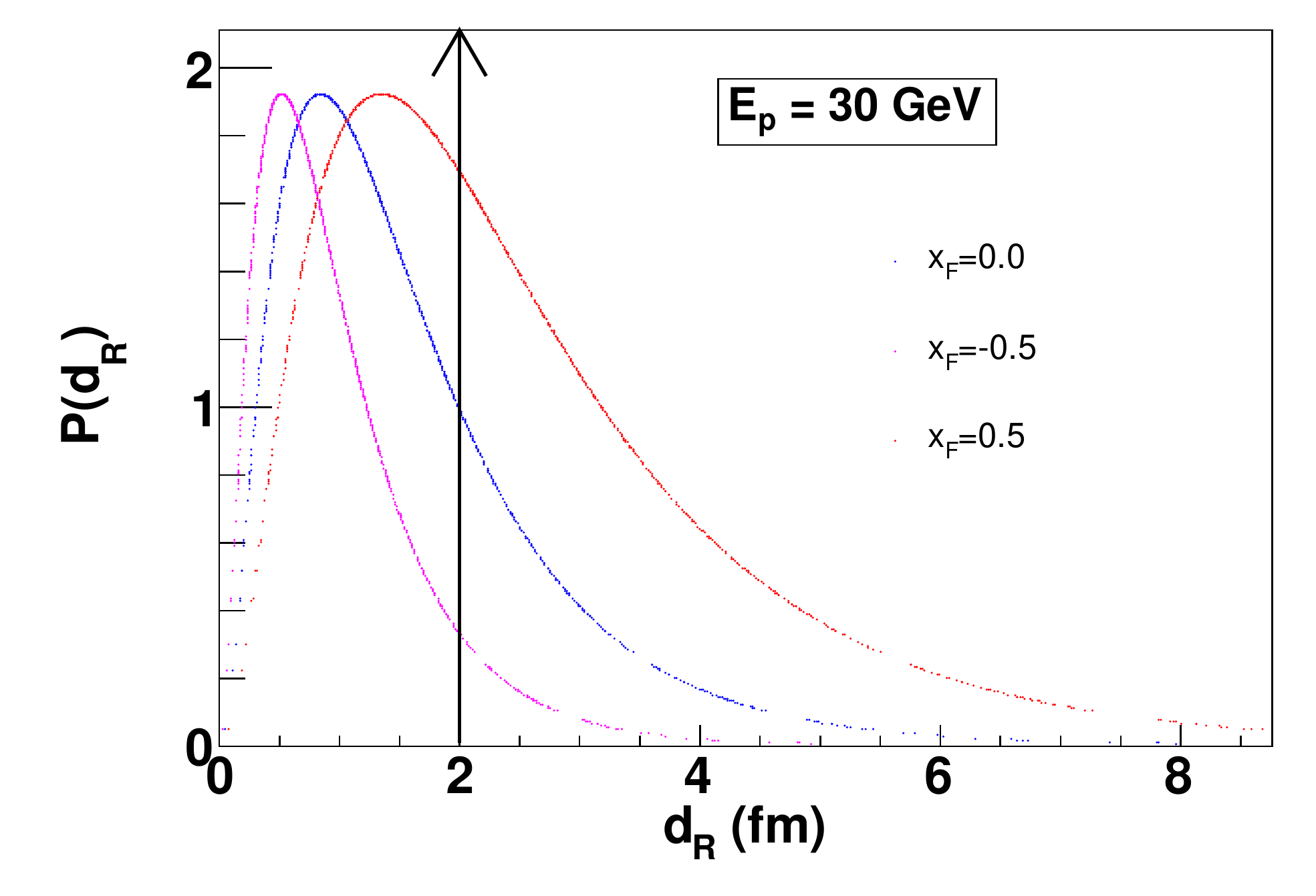}
}
\caption{Distribution of $J/\psi$ formation lengths in 15 (left panel) and 30 (right panel) GeV $p+A$ collisions at $x_F=-0.5$ (magenta), $0.0$ (blue) and $0.5$ (red).}\label{Fig8}
\end{figure*}
\subsubsection{Evaluation of survival probability}
\label{sec:6}
We now have all the ingredients needed to calculate the $J/\psi$ survival probability ($S_{J/\psi}$) at FAIR. Fig.~\ref{Fig5} displays the $J/\psi$ suppression pattern as a function of $x_F$ for $p+Au$ collisions of $15$ and $30$ GeV proton energy. For $15$ GeV, the constant geometric cross sections ($\sigma_{abs}^{Geo}$)  reduces $S_{J/\psi}$ by $20 \%$ and remains mostly independent of $x_F$. This is because the absorption depends only on the path length of the $J/\Psi$ in the frame of the target nucleus.  
For the dynamical cross sections, $S_{J/\psi}$ varies with $x_F$. However, the tiny cross sections obtained from the perturbative models ($\sigma_{abs}^{QCD}$) create a negligible absorption and $S_{J/\psi}$ remains nearly unity. The hadronic models ($\sigma_{abs}^{Had}$) predict a much larger suppression at both the energies. 
At 15 GeV, the suppression shows a slow increase with $x_F$ due to opening of the $D\bar{D}N$ channel, which grows with energy. 
At 30 GeV a step appears in the survival probability, around $x_F = 0.4$ fm,  which marks the boundary between resonance region and transition region according to our calculations. Use of different absorption profiles in these two regions lead to this abrupt jump in the resulting survival probability. 
%%%%%%%%%%%%%%%%%%%

\begin{figure*}[t]
\center
\resizebox{0.95\textwidth}{!}{%
\includegraphics{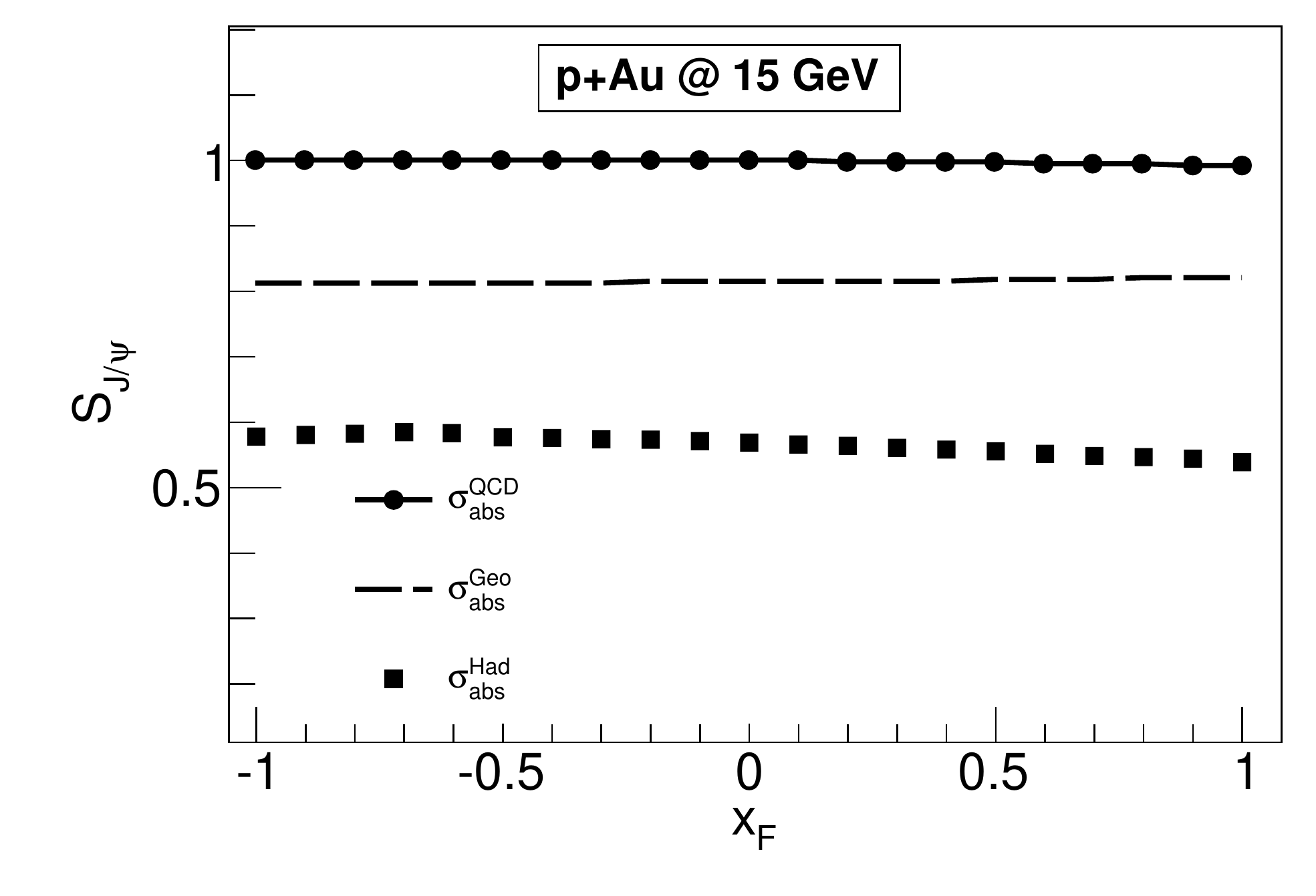}
\includegraphics{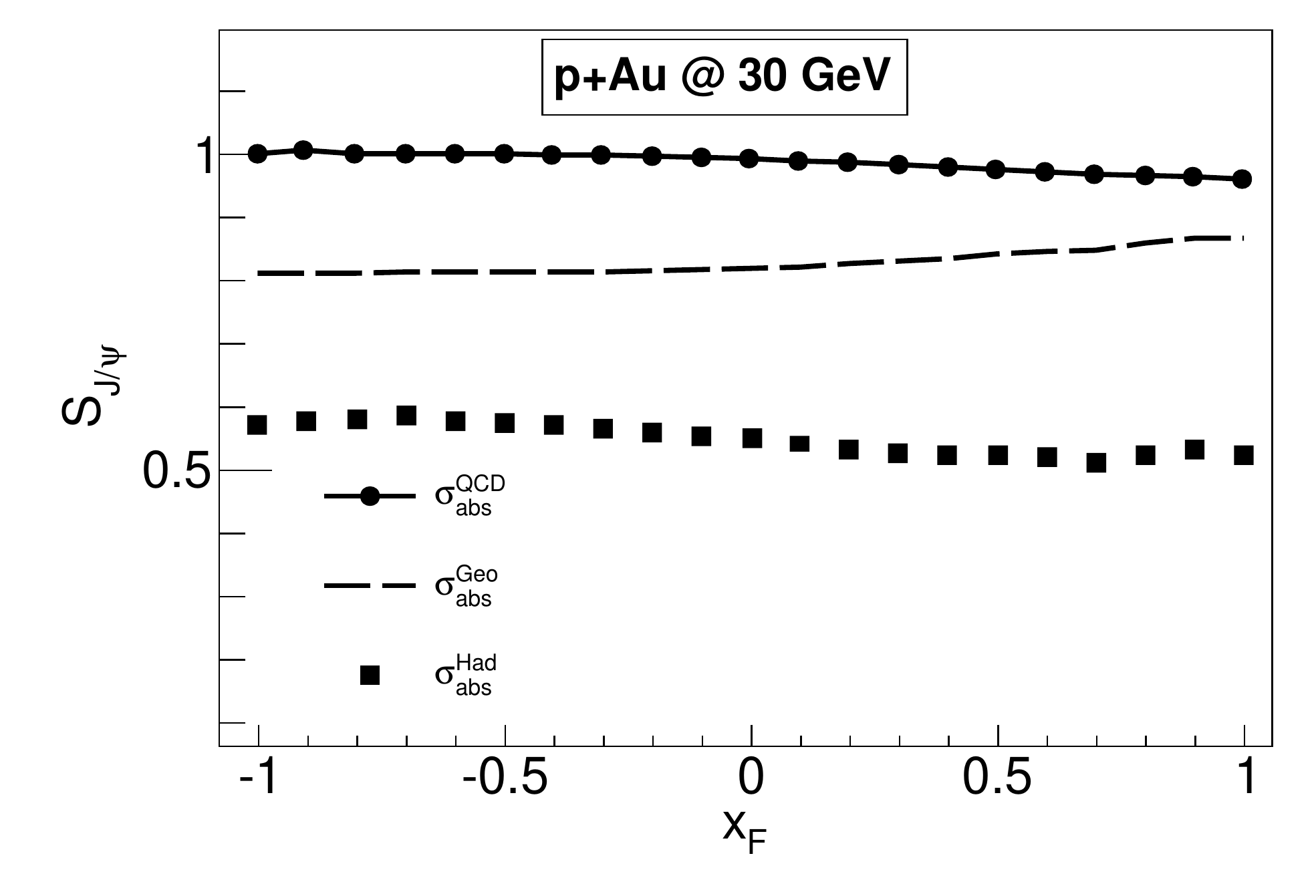}
}
\caption{$x_F$ dependence of $J/\psi$ survival probability within the variable formation time approach, for 15 and 30 GeV p+Au collisions.}\label{Fig9}
\end{figure*}
%%%%%%%%%%%%%%%%%%%%%%%%%%%%
Before we proceed further, it is interesting to test sensitivity of the $J/\psi$ suppression pattern on the different input parameters. Let us first investigate the impact of the resonance formation time ($\tau_{J/\psi}$) on the evaluation of $S_{J/\psi}$. For a given energy,  $\tau_{J/\psi}$ determines the fate of the produced $c\bar{c}$ pairs during their evolution inside the nuclear medium. So far, a constant value of $\tau_{J/\psi} \simeq 0.35$ fm is considered. But the choice of $\tau_{J/\psi}$ is not unique and highly model dependent. In~\cite{KharzeevThews}, for the first time, dispersion relations are used to reconstruct, in a model independent way, the formation dynamics of quarkonium states from the experimental data on $e^{+} e^{−} \rightarrow Q\bar{Q}$ annihilation. In contrast to a universal formation time, those calculations lead to a distribution of $\tau_{J/\psi}$ with a mean value of \mbox{$<\tau_{J/\psi}> = 0.44$ fm} and a width of $\delta_{J/\psi} = 0.31$ fm. This certainly calls for an investigation of the effect of such variable formation time to the observed survival probabilities. For this purpose we model the extracted distribution of $\tau_{J/\psi}$ ($P(\tau)$) as shown in Fig.~\ref{Fig7}. Hereafter, we generate the $J/\psi$ mesons randomly according to this distribution of $\tau_{J/\psi}$ for a given collision energy and $x_F$. The corresponding resonance formation lengths in the laboratory frame are given in Fig.~\ref{Fig8} for three typical $x_F$ values at backward, central and forward rapidities. At both the beam energies, most of the $J/\psi$ mesons are found to be formed within a spatial range of $2$ fm. For a given $x_F$, the survival probability now depends on the corresponding formation time. The average survival probability $<S(x_{F})>$ can then be obtained from the instantaneous survival probability ($S(x_{F}, \tau_{J/\psi}$) weighted with the distribution $P(\tau)$. The resulting suppression patterns are depicted in Fig.~\ref{Fig9}. In contrast to the situation of fixed $ \tau_{J/\psi}$, the sudden jump in the survival probability around $x_F \simeq 0.4$ in 30 GeV $p+Au$ collisions is now washed out due to averaging effects. For any $x_F$, the resonance region is more populated than the transition region. Otherwise the differences in $S_{J/\psi}$ between the two cases of fixed and variable $\tau_{J/\psi}$ are too meagre to be observed in the experimental data.   

\begin{figure}%[!h]
\center
\resizebox{0.5\textwidth}{!}{%
\includegraphics{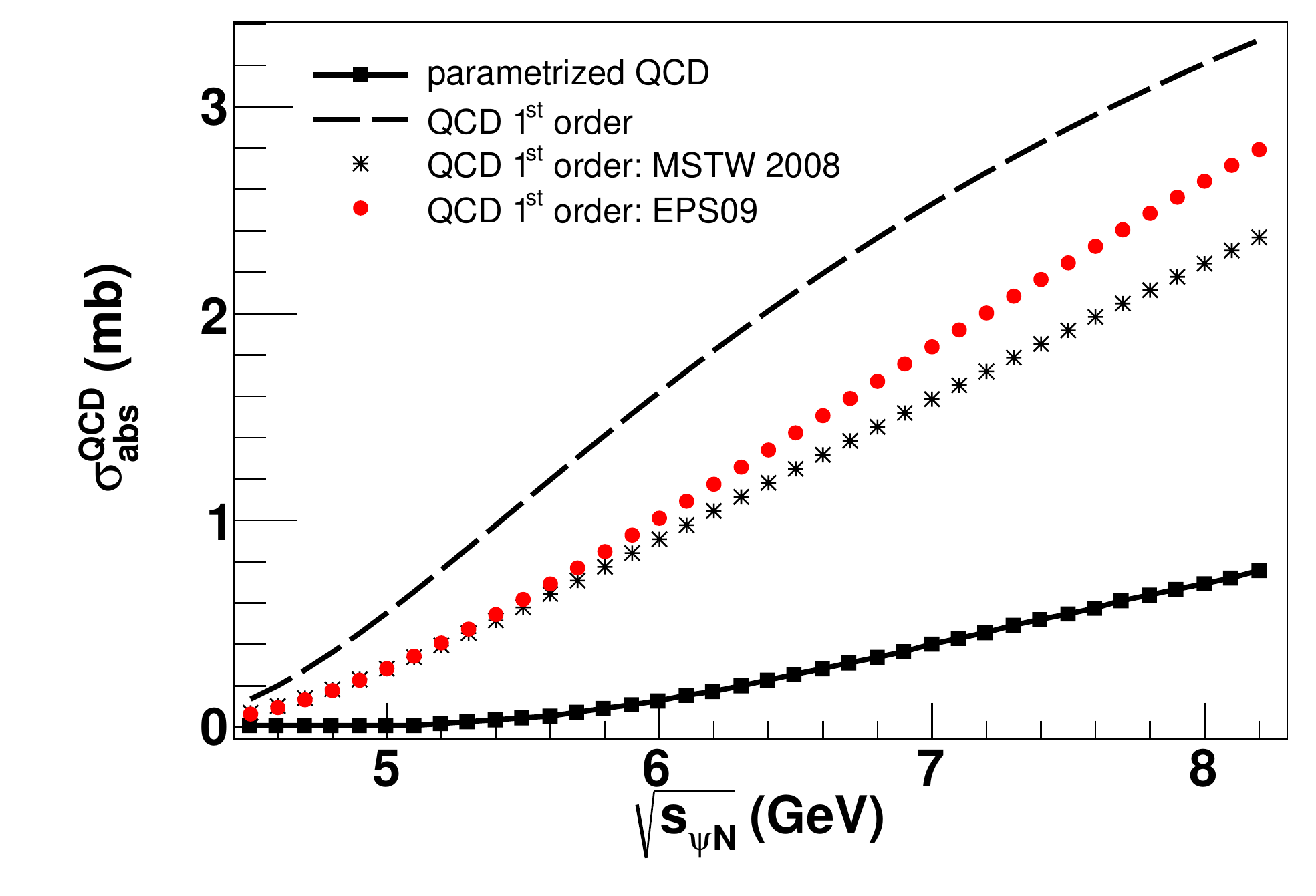}}
\caption{Comparison of $J/\psi +N$ dissociation cross sections from short distance QCD calculations for various parameterization of gluon density distributions.}\label{Fig11}
\end{figure}

\begin{figure*}[t]
\center
\resizebox{0.95\textwidth}{!}{%
\includegraphics{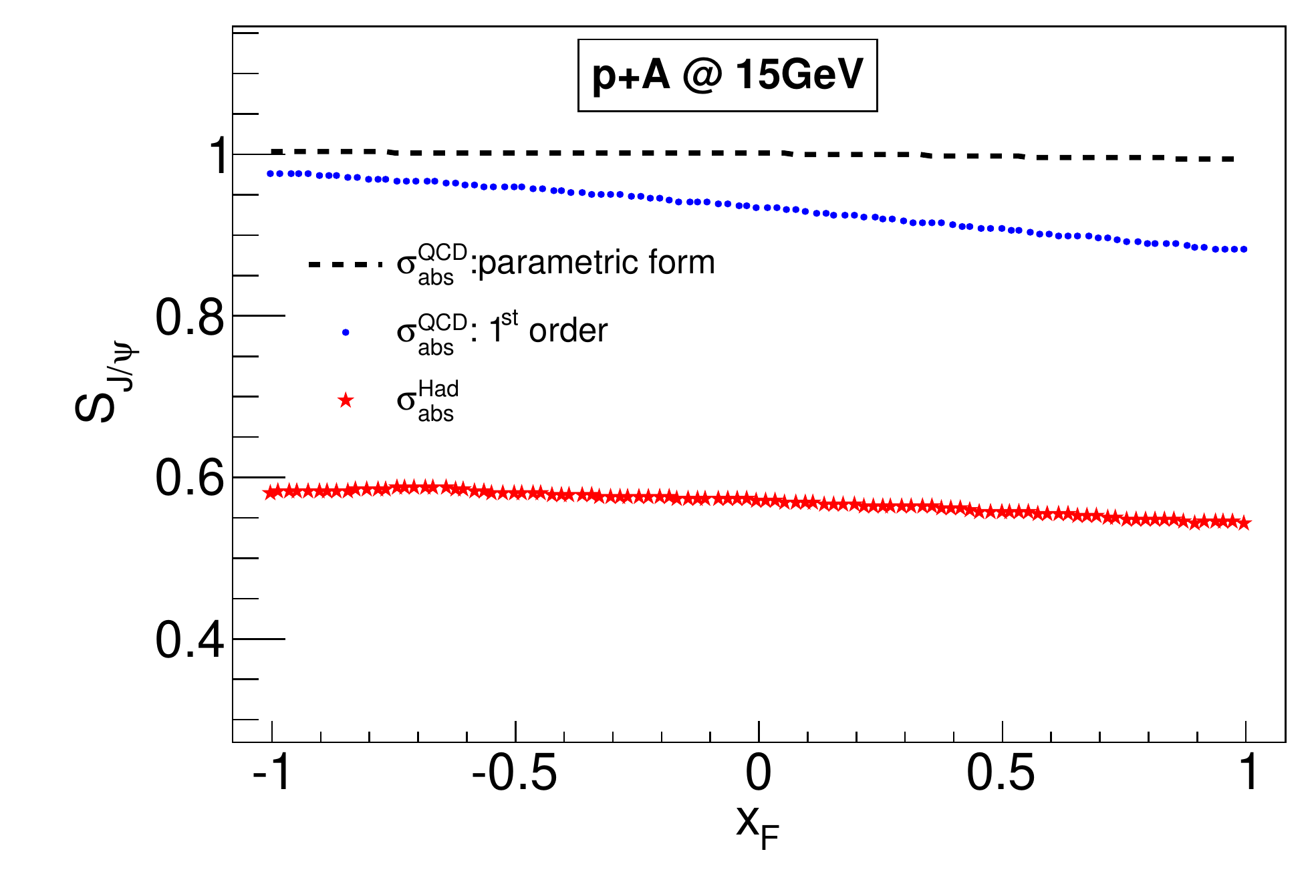}
\includegraphics{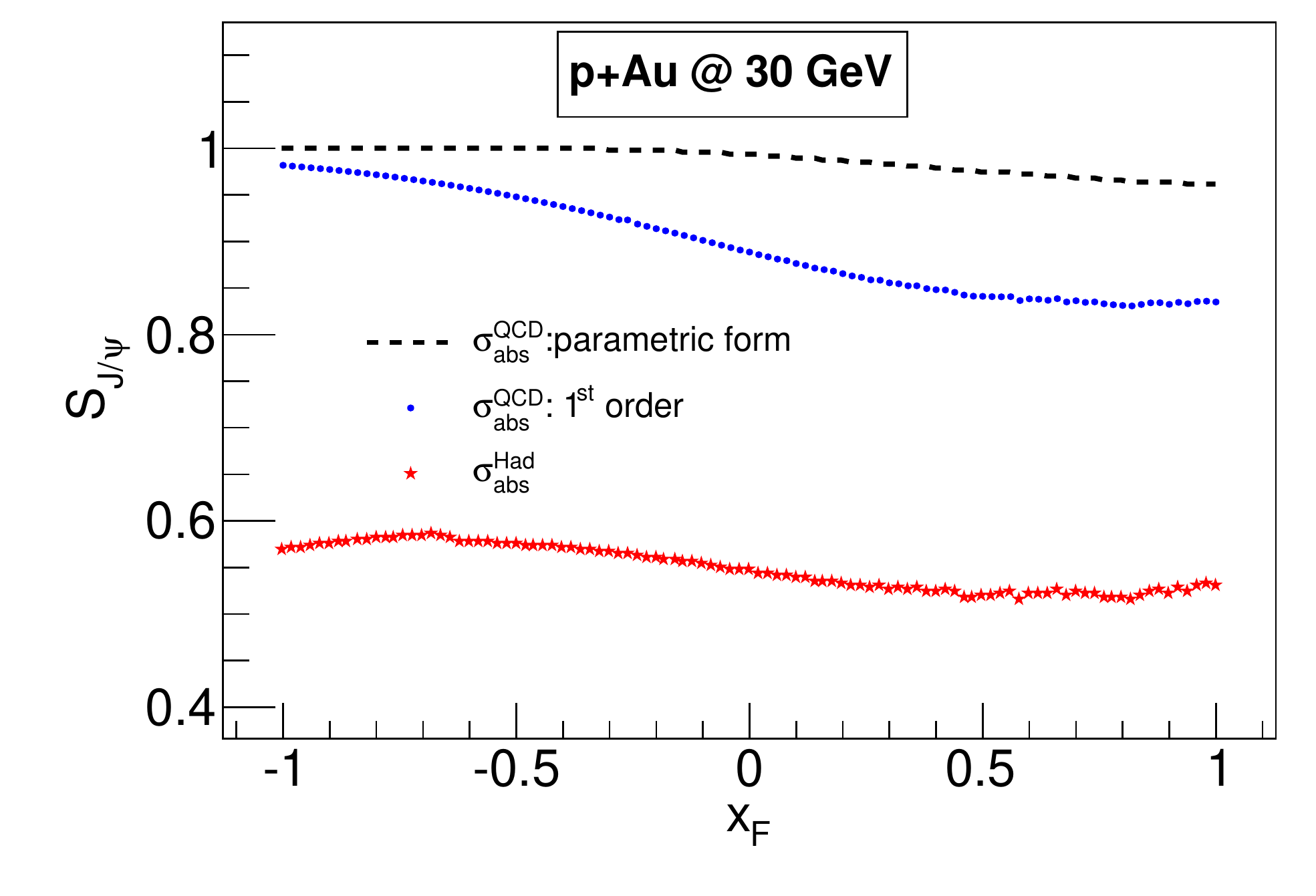}}
\caption{Comparison of $J/\psi$ survival probabilities for different estimations of QCD absorption cross sections. For completeness, the suppression from hadronic model calculation is also shown.}\label{Fig10}
\end{figure*}

Next we would like to explore the influence of the alternative estimations of $\sigma_{abs}^{QCD}$ on the resulting suppression pattern. In the previous calculations, $\sigma_{abs}^{QCD}$ has been extracted from the parametric form given in Eq.~\ref{Eq8}. Full QCD calculations of $J/\psi + N$ dissociation using short distance QCD methods based on operator product expansion are available in literature~\cite{KharzeevPLB99}. Within a first order approximation neglecting the correction terms due to finite mass of the nucleons, one obtains

\begin{equation}
\sigma_{abs}^{QCD}={{2^{13}\pi} \over {3^4\alpha_sm_c^2}} \int_{1 \over \xi}^{1} {{{{(\xi x - 1)}^{3 \over 2}} \over { {(\xi x)}^{5}}} {g(x) \over x} dx}
\label{Eq11}
\end{equation}
with $\xi={\lambda \over \epsilon_{J/\psi}}$, where $\lambda = {{s_{\psi N} - M_{J/\psi}^2 -m_{N}^2} \over 2M_{J/\psi}}$. The charm quark mass and the strong coupling constant are respectively denoted by $m_c$ and $\alpha_s$. $g(x)$ denotes the gluon distribution function in the nucleons inside the nucleus. We have evaluated $\sigma_{abs}^{QCD}$ within FAIR energy domain, for four different cases namely the parametric form given in Eq.~\ref{Eq8}, and estimations using Eq.~\ref{Eq11} with parameterization of gluon distribution $g(x)=2.5{(1-x)}^4$, and two more realistic gluon densities, namely MSTW 2008 leading order (LO) free proton parton distribution function (PDF)~\cite{MSTW} and EPS09 LO nuclear parton distribution function (nPDF)~\cite{EPS09}. Results are shown in Fig.~\ref{Fig11}. The first order QCD estimations with parameterized gluon distribution generates much larger dissociation compared to the parameterized dissociation cross section. The difference grows with increasing energy of the $J/\psi + N$ collisions. Results with the realistic gluon distributions lie in between but closer to the estimations with parameterized gluon distributions. 

The resulting effect of the amplified $\sigma_{abs}^{QCD}$ to the $J/\psi$ suppression scenario is shown in Fig.~\ref{Fig10} which includes the $J/\psi$ survival probability for two different estimations of $\sigma_{abs}^{QCD}$ producing smallest and largest absorptions. For meaningful comparison, the suppression within hadronic picture is also added. Calculations are done within the variable formation time approach as this possibly simulates the more realistic formation time dynamics. Maximum difference between two QCD estimates is $\simeq 13 \%$ at close to $x_F =1$. Otherwise the difference is minimal and a distinct separation from the hadronic case is still prominent.

\subsubsection{Predictions for $R_{pA}$}
\label{sec:7}
\begin{figure*}[t]
\center
\resizebox{0.95\textwidth}{!}{%
\includegraphics{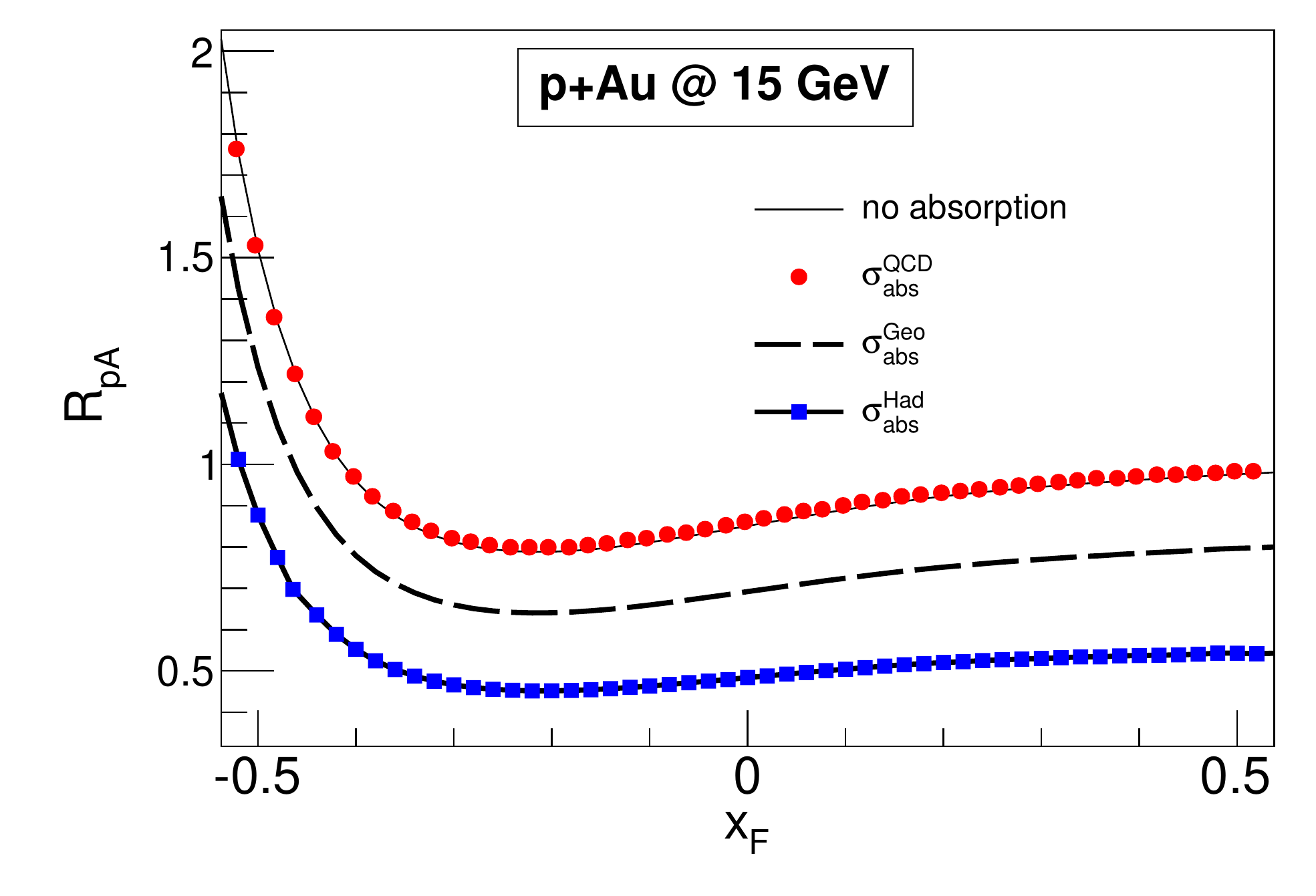}
\includegraphics{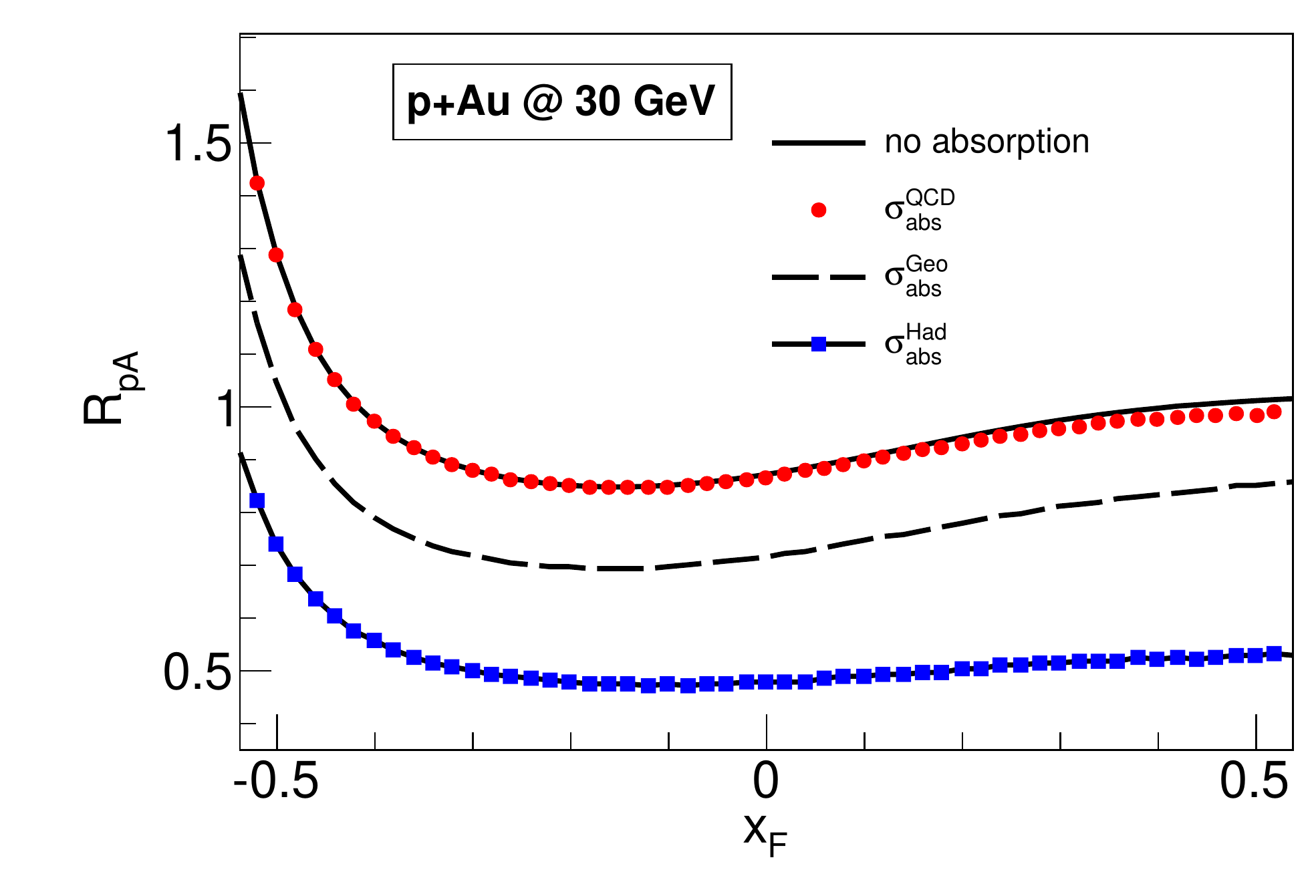}}
\caption{$x_F$ dependence of  $J/\psi$ $R_{pAu}$ in 15 and 30 GeV $p+Au$ collisions, with variable formation time approach.}\label{Fig6}
\end{figure*}
In experiments, the survival probability  $S_{J/\psi}$ is hard to measure as this requires to isolate the final state absorption of the particles from the initial state effects determining the particle production. Therefore, experimental collaborations report rather the $J/\psi$ production cross sections for different target nuclei. From those measured cross sections, one can construct the ratio $R_{pA}$ defined as ratio of the production cross section in $p+A$ to that in $p+p$ collisions. This ratio encodes all the possible CNM effects that modify the production during different stages of the $J/\psi$ evolution.  Here, we give predictions for the $x_F$ dependence of $J/\psi$ $R_{pA}$ in $p+Au$ collisions at FAIR, within variable formation time approach. Besides the final state dissociation discussed above, we include the initial state modification of the parton densities inside the target nucleus. This effect modifies the overall $c\bar{c}$ production in $p+A$ collisions.

%In particular it is well known by now that initial state modification of the parton densities inside the target nucleus would modify the rate of overall $c\bar{c}$ production which in turn affect the charmonium yield. We thus also give predictions for the $x_F$ dependence of $R_{pA}$ in $p+Au$ collisions at FAIR. 
The $J/\psi$ production is calculated using Color Evaporation Model (CEM)~\cite{CEM}. %Within CEM, a fixed fraction of the total number of $c\bar{c}$ pairs produced within the mass window $2m_{c} < m_{c\bar{c}} < 2m_{D}$ are assumed to produce a particular charmonium state. 
The total $c\bar{c}$ production cross sections in $p+p$ are estimated for two leading order partonic sub processes namely $gg$ fusion and $q\bar{q}$ annihilation, with MSTW 2008~\cite{MSTW} LO free proton PDF. Higher order corrections are accounted by a phenomenological $K$ factor. In case of $p+Au$ collisions, the shadowing effects inside the $Au$ nucleus, are incorporated using the EPS09~\cite{EPS09} LO nPDF set. For both parton densities, we have used the central sets having minimum uncertainties. It might be interesting to note here that at the kinematic domian probed by the charmonium production at SPS corresponded to an antishadowing region, where the parton densities in nuclei are enhanced with respect to those at free protons. However at FAIR energies, close to mid-rapidity nuclear parton densities are depleted leading to a reduction of overall $c\bar{c}$ production cross sections, even in the absence of final state dissociation (see Ref.~\cite{partha2,cortese} for more details). 
 
 The $x_F$ dependence of $R_{pAu}$ for $J/\psi$  in 15 and 30 GeV $p+Au$ collisions are shown in Fig.~\ref{Fig6}, for the three absorption scenarios discussed above. For completeness, we also added a curve without any final state dissociation.  The production cross sections are found to be negligible beyond $|x_{F}| > 0.5$, and thus ignored. %The shape of the $R_{pAu}$  is governed by the effects of initial state effects. 
As evident, the suppression curves for different absorption mechanisms are clearly distinguishable. Given the unprecedented beam intensities aimed at FAIR accelerators, the collected data are expected to suffer from much smaller statistical uncertainties. This would make an experimental distinction amongst different dissociation patterns feasible.

At this juncture, we would like to remind our readers that subsequent to the predictions in~\cite{SatzPaper1}, E866/NuSea Collaboration~\cite{E866} at Fermilab observed for the first time, substantial differences in suppression pattern between the $\psi'$ and $J/\psi$ in $p+A$ collisions in the backward hemisphere. %The suppression for the $\psi'$ is stronger than that for the $J/\psi$ for backward hemisphere close to $x_F$ near zero, but became comparable to that for the $J/\psi$ for $x_F > 0.6$. 
Subsequent measurements by NA50 Collaboration at SPS~\cite{NA50-400,NA50-450} were also in qualitative agreement with this observation. Stronger absorption of $\psi'$ compared to $J/\psi$ was seen while measuring inclusive charmonium production cross sections around mid-rapidity in $p+A$ collisions for a variety of nuclear targets. Assuming resonance formation within nuclear core, Fermilab data and preliminary NA50 data collected at 450 GeV, were explained using models that employ radial expansion of the color transparent tiny $c\bar{c}$ pair to the full resonance~\cite{Arleo,Gerland2,Spieles}. However the formation time as well as the inelastic reaction cross sections were generally kept as free parameters among the others which were fixed from the data and shadowing corrections at the initial stage of production were not taken into account. In~\cite{Frawley}, the absorption cross sections were extracted over a broad range of collision energies starting from $d+Au$ collisions at $\sqrt{s_{NN}} = 200$ GeV at RHIC, down to $p+A$ collisions at 158 GeV at SPS, within the same model~\cite{Arleo} of expanding $c\bar{c}$ pairs but with shadowing corrections explicitly taken into account. 

As mentioned earlier, at SIS 100 measurements of $\psi'$ would not seem to be feasible, and charmonium measurements would possibly be confined only to the study of $J/\psi$ mesons, which are always formed inside the target nucleus. At SIS 100 energies $J/\psi$ production would occur close to the kinematic thershold, where the difference between the dissociation cross sections obtained from perturbative and non-perturbative approaches show maximum difference, which is evident from the $x_F$ dependence of the absorption profile. This makes these upcoming measurements very relevant and unique compared to the existing studies.

\subsection{Discussion}
\label{sec:8}
Before we close, it is important to take a note on the limitations of our present analysis. We employ pQCD to calculate $J/\psi$ production cross sections, at SIS100 energies. This assumption is debatable, see Refs.~\cite{partha3,kiselev,steinheimer}, for alternative approaches of $J/\psi$ production at near threshold beam energies. However, the validity of QCD factorization,  in the near threshold quarkonium production can only be tested with data for FAIR energies. 

The feed down contribution from the decay of excited states ($\chi_c, \psi'$) to the overall $J/\psi$ production has not been taken into account assuming their rare occurrence due to higher kinematic threshold.%. 

In the evaluation of $R_{pA}$, the parton energy loss, which might affect the momentum distribution of the produced $c\bar{c}$ pairs, is not accounted for. The transverse momentum ($p_T$) of the $J/\psi$ mesons is artificially set to zero. At FAIR, the $J/\psi$ mesons are expected to be produced with a very small $p_T$, which justifies this approximation.  

For extracting the $J/\psi$ dissociation cross sections from the hadronic models, we extrapolated the results of~\cite{Oset} to higher energies and combined with calculations obtained from a different article~\cite{Sibirstev}. Combining parameters obtained from two different calculations as much as extrapolating bears its associated uncertainties, which had to be accepted due to a lack of better suited calculations. A consistent calculation within the same model framework including all known contributions to $J/\psi + N$ inelastic interaction at FAIR energies would be highly welcome. 

%We also ignored theoretical calculations predicting a medium induced modification of the spectral properties of the charmed hadrons~\cite{sourav,LeeFair}, as their effect on the charmonium production would possibly be irrelevant.
\section{Summary}
%Heavy quark systems are in general beleived to be an excellent tool to learn about the QCD dynamics at finite density.
 In this work, we estimated the $J/\psi$ suppression pattern in $p+Au$ collisions in the kinematic domain suitable for the experiments planned at FAIR. Our calculations suggest that the slow $J/\psi$ mesons produced in those low energy collisions will propagate through the nuclear matter as fully developed physical resonances.  

In $p+A$ interactions the final state absorption effects are limited to the target nucleons alone. The possibility of additional suppression due to co-moving secondary hadrons is drastically reduced in $p+A$ relative to $A+B$ collisions at these low energies. The experimental examination of the $x_F$ (or rapidity) dependence in $p+A$ reactions is a very promising test of $J/\psi$ propagation in baryonic matter, because of the very distinct behaviour of the corresponding survival probability. % Measurement of such differential distributions at SIS100 energies would be highly rewarding in terms of performing pioneering studies on the interaction of the physical resonances in the baryonic matter. 
%Existing theoretical calculations do not corroborate among themselves and predict quite different results for dissociation cross sections.
%Our results are given both in terms of the $J/\psi$ survival probability suffering dissociation due to final state nuclear absorption and $R_{pA}$ that additionally includes initial state modification of the parton densities of the bound nucleons inside the target. 

The world data collected so far on $J/\psi$ production from different experiments, can not shed much light on this issue. Most of the data are collected at higher energies, where propagation of the pre-resonance $c\bar{c}$ states through the nuclear matter dominates the experimentally explored phase space domain.  Hence neither a direct extrapolation of those results (dissociation cross sections) to lower energies nor an application of these dissociation mechanisms at higher energies seems to be viable.  At FAIR energies, the various scenarios of $J/\psi$ absorption give a very distinct survival probability leading to distinguishably different pattern of $R_{pA}$. 
 
%Inclusion of random formation dynamics in contrast to the universal formation time approach does not bring any appreciable change on the overall survival pattern of $J/\psi$. 
Such measurements are certainly feasible with the CBM detector set up at SIS100 energies which would also be capable to experimentally discriminate among various absorption scenarios. These results will also prove useful for interpretation of heavy-ion data to be collected in future.

\section{Acknowledgements}
We would like to thank D. Kharzeev. E. Oset, S. Sarkar and J. Sadhukhan for many useful discussions. The work was done during the post-doctoral tenure of PPB at CBM department, GSI, Germany.
%PPB is grateful to the organizers of ``International Conference on Matter under High Densities'', Sikkim, India, for giving the opportunity to present an initial version of this work. %Comments and suggestions received there have certainly given the manuscript a better shape.
 Work of MD has been supported by the BMBF (06FY9099I and 05P12RFFC7), HIC for FAIR and GSI.


\begin{thebibliography}{100}
\bibitem{MS} T. Matsui and H. Satz, Phys. Lett. B  {\bf 178}, (1986) 416.

\bibitem{Vogt} R. Vogt, Physics Reports 310, 197 (1999).
\bibitem{Satz} H. Satz, J. Phys. G32:R25,2006; H. Satz, Rept. Prog. Phys.63:1511,2000; 
\bibitem{Kluberg} L. Kluberg and H. Satz, arXiv:hep-ph/0901.3831.

\bibitem{NA3} J. Badier {\it et al.}, Z. Phys. C{\bf 20}, 101 (1983).

\bibitem{E772} D. Alde {\it et al.}, Phys. Rev. Lett. {\bf 66}, 133 (1991).

\bibitem{NA38} The NA38 Collaboration, C. Baglin et al., Phys. Lett. B{\bf 220}, 471 (1989); B{\bf 221}, 465 (1990); B{\bf 221}, 472 (1990); B{\bf 225}, 459 (1991).

\bibitem{NA50-200} M.C. Abreu {\it et al.}, NA50 Collaboration, Phys. Lett. B{\bf 410}, 337 (1997)

\bibitem{NA50-400} B. Alessandro {\it et al.} NA50 Collaboration, Euro. J.Phys {\bf 48} 329 (2006).


\bibitem{NA50-450} B. Alessandro {\it et al.} NA50 Collaboration, Euro. J.Phys {\bf 33} 31 (2004).

\bibitem{E866} M. J. Leitch {\it et al.} E866 Collaboration, Phys. Rev. Lett. {\bf 84} 3256 (2000).

\bibitem{HERAB} I. Abt {\it et al.} HERA-B Collaboration, Eur. Phys. J. C {\bf 60} 525 (2009) 

\bibitem{NA60} Roberta Arnaldi, for the NA60 Collaboration, Nucl. Phys. A{\bf 830} 345c-352c, (2009); R. Arnaldi {\it et. al.}, NA60 Collaboration, Phys. Lett. B {\bf 706} 263 (2012).

%\bibitem{Ale05} B.~Alessandro {\it et al.} NA50 Collaboration,  Eur. Phys. J. {\bf C39} (2005) 335.

%\bibitem{NA50-400} B. Alessandro {\it et al.} NA50 Collaboration, Euro. J.Phys {\bf 48} (2006) 329.

\bibitem{Khar-Glauber} D. Kharzeev {\it et al.}, Z. Phys. C {\bf 74} (1997) 307.

\bibitem{Lourenco} C. Lourenco, R. Vogt and H. K. Wohri, JHEP {\bf 0902} (2009) 014.

\bibitem{SatzPaper1} D. Kharzeev and H. Satz, Phys. Lett. B {\bf 356} (1995) 365.

\bibitem{cbm-paper} T. Ablyazimov {\it et al.} CBM Collaboration, Eur. Phys. J. A {\bf 53} (2017) 60.

\bibitem{cbm-satz} H. Satz, Talk given at the ``HICforFAIR workshop: Heavy flavor physics with CBM'' (https://indico.gsi.de/conferenceDisplay.py?confId=2474)

\bibitem{Anton} A. Andronic {\it et al.}, Eur. Phys. J. C{\bf 76} (2016) 107.


\bibitem{Karsch} F. Karsch, H. Satz, Z. Phys. C {\bf 51} (1991) 209.


\bibitem{formation} R.L. Thews, Nucl. Phys. (Proc. Supp.) {\bf 23B} (1991) 362;
J.P. Blaizot, R. Venugopalan, M. Prakash, Phys. Rev. D {\bf 45} (1992) 814;
J. Hufner and B.Z. Kopeliovich, Phys. Lett. B {\bf 426} (1998) 154.


\bibitem{Wongbook} C. Y. Wong {\it Introduction to High-Energy Heavy-Ion Collisions}, World Scientific Publisher, 1994.


%\bibitem{Gerland} G. Farrar, L. Frankfurt, M. Strikman, H. Li,  Phys. Rev. Lett. {\bf 64}, (1990) 2996; L. Gerland {\it et. al.}, Phys. Rev. Lett. {\bf 81} (1998) 762, arXiv:nucl-th/9803034.

%\bibitem{Ring} P. Ring and P. Schuk, {\it The Nuclear Many-Body Problem}, Springer Publisher ISSN 0172-5998.

%\bibitem{Gerland1} L. Gerland {\it et. al.}, Nucl. Phys. A {\bf 663} (2000) 1019.
%\bibitem{Gerland2} L. Gerland, L. Frankfurt, M. Strickman and H. Stocker, Phys. Rev. C {\bf 69} (2004) 014904; arXiv:nucl-th/0307064.

%\bibitem{Dosch} H. Dosch, F. S. navarra, M. Nielsen and M. Rueter, Phys. Lett. B {\bf 466} (1999) 363; arXiv:hep-ph/9908274.

%\bibitem{Hufner1987} J. Hufner and B. Povh, Phys. Rev. Lett. {\bf 58} (1987) 1612.


\bibitem{Arleo} F. Arleo, P.B. Gossiaux, T. Gousset and J. Aichelin, Phys. Rev. C {\bf 61} 054906 (2000); hep-ph/9907286.


\bibitem{KharzeevPLB94} D. Kharzeev and H. Satz, Phys. Lett. B {\bf 334} (1994) 155.


\bibitem{KharzeevPLB99} D. Kharzeev and H. Satz, Phys. Lett. B {\bf 389} (1996) 595.

\bibitem{SHLee} Yongseok Oh, Sungsik Kim, and Su Houng Lee, Phys. Rev. C {\bf 65} (2002) 067901, arXiv:hep-ph/0111132; Taesoo Song and Su Houng Lee, Phys. Rev. D {\bf 72} (2005) 034002.


\bibitem{Redlich} K. Redlich, H. Satz and G. M. Zinovjev, Eur. Phys. J. C {\bf 17} (2000) 461.


\bibitem{Martins} K. Martins, D. Blaschke and E. Quack, Phys. Rev. C {\bf 51} (1995) 2723.

\bibitem{Brane} J. P. Hilbert, N. Black, T. Barnes, and E. S. Swanson, Phys. Rev. C {\bf 75} (2007) 064907, nucl-th/0701087.

\bibitem{Muller} S. G. Matinyan and B. Mueller, Phys. Rev. C {\bf 58} (1998) 2994.

\bibitem{Haglin} Kevin L. Haglin, Phys. Rev. C {\bf 61} (2000) 031902 (R); nucl-th/9907034.


\bibitem{Oset} R. Molina, C. W. Xiao and E. Oset, Phys. Rev. C {\bf 86} (2012) 014604; nucl-th/12030979


%\bibitem{CMKo} Yongseok Oh, Taesoo Song, Su Houng Lee, and Cheuk-Yin Wong, J.Korean Phys.Soc. {\bf 43} (2003) 1003, arXiv:nucl-th/0205065; Yongseok Oh, Wei Lu and C. M. Ko, Phys. Rev. C {\bf 75} (2007) 064903, arXiv:nucl-th/0702077 

\bibitem{Sibirstev} A. Sibirstev, K. Tsushima and A. W. Thomas, Phys. Rev. C {\bf 63} (2001) 004906.

\bibitem{KharzeevThews} D. Kharzeev and R. L.Thews, Phys. Rev. C {\bf 60} (1999) 041901; arXiv:nucl-th/9907021.


\bibitem{MSTW} A. D.  Martin, W. J. Stirling, R. S. Thorne, G. Watt, Eur. Phys. J. C {\bf 63} (2009) 189; A. D. Martin, W. J. Stirling, R. S. Thorne, G. Watt, Eur. Phys. J. C {\bf 64} (2009) 653; A. D. Martin, W. J. Stirling, R. S. Thorne, G. Watt, Eur. Phys. J. C {\bf 70} (2010) 51.

\bibitem{EPS09} K.J. Eskola, H. Paukkunen and C.A. Salgado, JHEP {\bf 04} (2009) 065.

\bibitem{CEM}  M.B. Einhorn, S.D. Ellis, Phys.  Rev.  D {\bf 12} (1975) 2007; H. Fritzsch, Phys. Lett. B {\bf 67} (1977); M. Glueck, J.F. Owens, E. Reya, Phys. Rev. D {\bf 17} (1978) 2324; J. Babcock, D. Sivers, S. Wolfram, Phys. Rev. D {\bf 18} (1978) 162.

\bibitem{partha2}  P. P. Bhaduri, A. K. Chaudhuri and S. Chattopadhyay, Phys. Rev.C {\bf 89},044912 (2014). 

\bibitem{cortese} R. Arnaldi, P. Cortese and E. Scomparin, Phys. Rev. C {\bf 81}, 014903 (2010). 

\bibitem{Gerland2} L. Gerland, L. Frankfurt, M. Strickman and H. Stocker, Phys. Rev. C {\bf 69} (2004) 014904; arXiv:nucl-th/0307064.


\bibitem{Spieles} C. Spieles {\it et. al.}  Phys.Rev. C {\bf 60} (1999) 054901; hep-ph/9902337.


\bibitem{Frawley} D. C. McGlinchey, A. D. Frawley and R. Vogt, Phys. Rev. C {\bf 87}, 054910 (2013); arXiv:1208.2667 [nucl-th].

\bibitem{partha3} P. P. Bhaduri and S. Gupta, Phys. Rev. C {\bf 88}, 045205 (2013).

\bibitem{kiselev}
  Y.~T.~Kiselev, E.~Y.~Paryev and Y.~M.~Zaitsev,
  %``Near-threshold J/Ψ production in proton–nucleus collisions,''
  Int.\ J.\ Mod.\ Phys.\ E {\bf 23} (2014) no.12, 1450085
  doi:10.1142/S0218301314500852
  [arXiv:1409.2428 [nucl-th]].

\bibitem{steinheimer} 
  J.~Steinheimer, A.~Botvina and M.~Bleicher,
  %``Sub-threshold charm production in nuclear collisions,''
  Phys.\ Rev.\ C {\bf 95}, no. 1, 014911 (2017)
  doi:10.1103/PhysRevC.95.014911
  [arXiv:1605.03439 [nucl-th]].

\end{thebibliography}
\end{document}